\documentclass[twocolumn,showpacs,floatfix,prd,amsmath,amssymb]{revtex4}
\usepackage{graphicx}

\newcommand{\del}[1]{ \partial_{#1} }

\def\al{{\alpha}}
\def\lam{{\lambda}}
\def\gam{{\gamma}}

\def\tilU{{\tilde{U}}}
\def\tU{{\tilde{U}}}
\def\mE{{{\mathcal E}}}
\def\at{{\tilde{\alpha}}}
\def\pt{{\tilde{p}}}
\def\qt{{\tilde{q}}}

\begin{document}
\title{
Solitonic generation of vacuum solutions 
in five-dimensional General Relativity
}
\author{Hideo Iguchi and Takashi Mishima} 
\affiliation{
Laboratory of Physics,~College of Science and Technology,~
Nihon University,\\ Narashinodai,~Funabashi,~Chiba 274-8501,~Japan
}
\date{\today}
\begin{abstract}
We describe a
solitonic solution-generating technique for the five-dimensional 
General Relativity. Reducing the five-dimensional problem to 
the four-dimensional one, we can systematically obtain
single-rotational axially symmetric vacuum solutions. 
Applying the technique for a simple 
seed solution, we have previously obtained the series of stationary solutions 
which includes $S^2$-rotating black ring.
We analyze the qualitative features
of these solutions, e.g., conical singularities, closed timelike curves, and
spacetime curvatures. 
We investigate the rod structures of seed and solitonic solutions.
We examine the relation between the expressions of the metric in the prolate-spheroidal coordinates and
in the C-metric coordinates.
\end{abstract}      
\maketitle

\section{Introduction}
In the recent years, finding exact solutions of 
higher-dimensional General Relativity has attracted much interest.
There are several reasons for this.
The string theory, which is a promising candidate of a quantum gravity theory,
predicts that the spacetime has more than four dimensions.  
Furthermore the possibilities of the large or infinite extradimensions 
are proposed for solving the hierarchy problem
\cite{{Arkani-Hamed:1998rs},{Randall:1999ee}}.
Also, the production of higher-dimensional black holes in future
linear colliders is predicted based on these models  \cite{Giddings:2001bu}.

Although the uniqueness theorem has not been
generalized to the higher dimensions yet,
the studies of 
the spacetime structures in higher-dimensional 
General Relativity revealing 
the rich structure have been performed recently with great intensity. 
For example,
several authors examined some qualitative 
features concerning the black hole horizon topologies 
in higher dimensions \cite{ref0}.
This possibility of 
the variety of horizon topologies gives difficulty 
to the establishment of theorems 
analogous with the powerful uniqueness theorem 
in four dimensions.
Also several exact solutions involving black holes were 
obtained in the higher-dimensional spacetime.
The higher-dimensional generalization of 
the four-dimensional exact solutions were obtained for cases of the
Schwarzschild and Reisner-Nordstr\"{o}m black holes by Tangherlini \cite{Tangherlini:1963}
and for the Kerr black hole by Myers and Perry \cite{refMP}. 
Particularly in the five-dimensional case,
several researchers have tried to search new exact solutions
since the remarkable discovery of a rotating black ring solution
with horizon topology
$S^1 \times S^2$ by Emparan and Reall \cite{ref1}. 
For example, the supersymmetric black rings \cite{ref2} 
and the black ring solutions under the influence of external 
fields \cite{ref3} are found.
The systematical derivation of these solutions and some generalizations
were examined by Yazadjiev \cite{Yazadjiev:2005hr}.
In addition the richness of the phase structure of Kaluza-Klein
black holes have been discussed. 
Above all, the phases with and without Kaluza-Klein bubbles
\cite{{Witten:1981gj},{Elvang:2002br}} have been investigated
energetically.
(See \cite{{Elvang:2004ny},{Elvang:2004iz}} and references therein.)
Also the phase transition between black holes and black strings 
are widely investigated. (See, e.g., \cite{Kol:2004ww} and references therein.)
New geometrical structure of charged static black hole in the 
five-dimensional Einstein-Maxwell theory has been studied \cite{Ishihara:2005dp}.

In this context
a systematical search of possible solutions in higher dimensions　
is of great significance.
In the four-dimensional General Relativity, 
the solution-generating techniques had been fully developed 
for the stationary and axisymmetric spacetime.
These techniques were used for the systematical 
generations of 
solutions \cite{Stephani:2003tm},
including the famous multi-Kerr solutions \cite{ref6}, 
since the discovery of Tomimatsu-Sato solutions \cite{ref5}.
As in the four-dimensional case, 
the development of systematical ways of constructing new solutions 
would promote our understanding of the higher-dimensional General Relativity.  

Using the fact that finding some class of five-dimensional solutions 
can be reduced to the four-dimensional problem \cite{{Dereli:1977},{Bruckman:1985yk},{Mazur:1987}},
the present authors \cite{Mishima:2005id} obtained a new class of
five-dimensional stationary solutions by using 
a kind of B\"{a}cklund transformation.
In the analysis the formula given by Castejon-Amenedo
and Manko  \cite{ref7} were applied. 
(See also \cite{Gutsunaev:1988xh}.)
This method is promising in the point that we can easily
perspect the property of the solution and
obtain the exact expression of it.
It was shown that the solutions obtained in \cite{Mishima:2005id} include 
a single-rotational black ring solution with $S^1 \times S^2$
event horizon topology 
which rotates in the azimuthal direction of $S^2$.
This $S^2$-rotating black ring solution is important, especially when we 
try to find
a double-rotational black
ring solution, because it should be realized when we take
a single-rotational limit of the
double-rotational black ring.
After the discovery of the solution,
Figueras found a C-metric expression of $S^2$-rotating black ring solution
\cite{Figueras:2005zp}.
Tomizawa et al. \cite{Tomizawa:2005wv}
showed that the same black ring solution
is obtained by using the 
inverse scattering method \cite{refBZ}, which has potential to 
produce more general solutions.
This techniques was applied to the higher-dimensional theory of
Kaluza-Klein compactifications several decades ago \cite{refKK}. 
Recently it was applied to the
five-dimensional static  Einstein equation \cite{Koikawa:2005ia}.
Also this technique was used to rederive the five-dimensional Myers and Perry solution \cite{Pomeransky:2005sj}.
Using this technique and a matrix transformation of Ehlers type,
solitonic solutions of 
five-dimensional string theory system were obtained \cite{{Herrera-Aguilar:2003ui},{Herrera-Aguilar:2005sa}}.

In this paper we present a detailed explanation for the solution-generating
technique used to derive the new solutions
in the previous paper \cite{Mishima:2005id}. 
In addition we analyze the qualitative features of the solutions
including singular ones in detail.
These solutions are generated from
the five-dimensional Minkowski spacetime as a seed solution.
Although the spacetimes found here have singular objects like 
closed timelike curves (CTC) and naked curvature singularities 
in general, a part of these solutions 
is a new class of black ring solutions whose 
rotational planes are different from those of Emparan and Reall's.
This black ring solution needs a conical singularity inside or 
outside the ring because the effect of rotation cannot compensate 
for the gravitational attractive force.
The excess angle of the ring can be represented by mass and radius parameters.
When we fix the mass and radius parameters, there is an upper limit of 
the rotational parameter of the ring.

We also study the rod structures \cite{{ref8},{refHAR}} of the seed solution and the corresponding 
solitonic solution 
to understand the relation between them.
The rod structure analysis help us to find the
seed solution for the solitonic solution we want to obtain. 
Recently the seed solution of Emparan and Reall's $S^1$-rotating
black ring has been obtained following this strategy \cite{Iguchi:2006rd}.
It has been reported that the $S^1$-rotating black ring solution
is generated starting from the Levi-Civita metric by using 
the inverse scattering method \cite{Tomizawa:2006vp}.

Particularly in the $S^2$-rotating black ring,
the C-metric representation of the solution
was obtained  \cite{Figueras:2005zp}.
This representation is simple and suitable for 
the analysis of the solutions.
In this paper we 
investigate the relation
between these two expressions in the prolate-spheroidal and C-metric coordinates. 

The plan of the paper is as follows.  In Sec. \ref{sec:technique}
we describe the solution-generating technique used in this analysis.
In Sec. \ref{sec:application} we denote the application of the technique for the simple seed solution. We analyze the properties of the solutions derived 
in the application in Sec. \ref{sec:property}. 
In Sec. \ref{sec:summary} we give a summary of this article. 
\section{Solitonic solution-generating technique}
\label{sec:technique}
In this section we write up the procedure for generating the axisymmetric solution
in the five-dimensional General Relativity,
which was applied to the generation 
of the single-rotational black ring solution whose rotational direction
is different from the one of Emparan-Reall's ring \cite{Mishima:2005id}. 
In this approach we use the fact that the five-dimensional problem with some
conditions can be reduced to a four-dimensional problem, 
which had been investigated elaborately several decade ago.   
As a result, we can use several well-established solution-generating
techniques in the four dimensions to generate new five-dimensional solutions
in a straightforward way.
Also if we prepare various seed solutions for each applications, then we can obtain
various kinds of new solutions even within the limitation of single-rotation.

The spacetimes which we considered 
satisfy the following conditions:
(c1) five dimensions, (c2) asymptotically flat spacetimes, 
(c3) the solutions of 
vacuum Einstein equations, (c4) having three commuting Killing vectors 
including time translational invariance and 
(c5) having a single nonzero angular momentum component. 
Under the conditions (c1) -- (c5), 
we can employ the following Weyl-Papapetrou metric form 
(for example, see the treatment in \cite{refHAR}), 
\begin{eqnarray}
ds^2 &=&-e^{2U_0}(dx^0-\omega d\phi)^2+e^{2U_1}\rho^2(d\phi)^2
       +e^{2U_2}(d\psi)^2 \nonumber \\ && 
       +e^{2(\gamma+U_1)}\left(d\rho^2+dz^2\right) ,
       \label{WPmetric}
\end{eqnarray}
where $U_0$, $U_1$, $U_2$, $\omega$ and $\gamma$ are functions of 
$\rho$ and $z$. 
Then we introduce new functions 
$S:=2U_0+U_2$ and $T:=U_2$ so that 
the metric form (1) is rewritten into 
\begin{eqnarray}
ds^2 &=&e^{-T}\left[
       -e^{S}(dx^0-\omega d\phi)^2
       +e^{T+2U_1}\rho^2(d\phi)^2 \right. \nonumber \\&&\hskip 0cm \left.
+e^{2(\gamma+U_1)+T}\left(d\rho^2+dz^2\right) \right]
  +e^{2T}(d\psi)^2.
  \label{MBmetric}
\end{eqnarray}
Using this metric form
the Einstein equations are reduced to the following set of equations, 
\begin{eqnarray*}
&&{\bf\rm (i)}\quad
\nabla^2T\, =\, 0,   \\
&&{\bf\rm (ii)}\quad
\left\{\begin{array}{ll}
& \del{\rho}\gamma_T={\displaystyle
  \frac{3}{4}\,\rho\,
  \left[\,(\del{\rho}T)^2-(\del{z}T)^2\,\right]}\,\ \   \\[3mm]
& \del{z}\gamma_T={\displaystyle 
\frac{3}{2}\,\rho\,
  \left[\,\del{\rho}T\,\del{z}T\,\right],  }
 \end{array}\right.  \\
&&{\bf\rm (iii)}\quad
\nabla^2\mE_S=\frac{2}{\mE_S+{\bar\mE}_S}\,
                    \nabla\mE_S\cdot\nabla\mE_S , \\  
&&{\bf\rm (iv)}\quad
\left\{\begin{array}{ll}
& \del{\rho}\gamma_S={\displaystyle
\frac{\rho}{2(\mE_S+{\bar\mE}_S)}\,
  \left(\,\del{\rho}\mE_S\del{\rho}{\bar\mE}_S
  -\del{z}\mE_S\del{z}{\bar\mE}_S\,
\right)}     \\
& \del{z}\gamma_S={\displaystyle
\frac{\rho}{2(\mE_S+{\bar\mE}_S)}\,
  \left(\,\del{\rho}\mE_S\del{z}{\bar\mE}_S
  +\del{\rho}\mE_S\del{z}{\bar\mE}_S\,
  \right)},  
\end{array}\right.  \\
&&{\bf\rm (v)}\quad
\left( \del{\rho}\Phi,\,\del{z}\Phi \right)
=\rho^{-1}e^{2S}\left( -\del{z}\omega,\,\del{\rho}\omega \right),  \\
&&{\bf\rm (vi)}\quad 
\gamma=\gamma_S+\gamma_T,   \\
&&{\bf\rm (vii)}\quad 
U_1=-\frac{S+T}{2},
\end{eqnarray*}
where $\Phi$ is defined through the equation (v) and the function 
$\mathcal{E_S}$ is defined by 
$
\,\mE_S:=e^{S}+i\,\Phi\,.
$
The equation (iii) is exactly the same as the Ernst equation in four dimensions \cite{refERNST}, 
so that we can call $\mE_S$ the Ernst potential. 
The most nontrivial task to obtain new metrics is to solve 
the equation (iii) because of its nonlinearity. 
To overcome this difficulty we can however use the methods already 
established in the four-dimensional case. 
Here we use the method similar to the Neugebauer's 
B\"{a}cklund transformation \cite{Neugebauer:1980} 
or the HKX transformation \cite{Hoenselaers:1979mk}, whose essential idea 
is that new solutions are generated by adding solitons to seed spacetimes. 

To write down the exact form of the metric functions,
we follow the procedure
given by Castejon-Amenedo and Manko \cite{ref7},
in which they 
discussed a deformation of a Kerr black hole 
under the influence of some external gravitational fields.
In the five-dimensional space time we start from the following form of a 
static seed metric
\begin{eqnarray}
ds^2 &=& e^{-T^{(0)}}\left[
       -e^{S^{(0)}}(dx^0)^2
       +e^{-S^{(0)}}\rho^2(d\phi)^2  \right. \nonumber \\ && \left.
   +e^{2\gamma^{(0)}-S^{(0)}}\left(d\rho^2+dz^2\right) \right]
  +e^{2T^{(0)}}(d\psi)^2.
\end{eqnarray}
For this static seed solution, $e^{S^{(0)}}$, of the Ernst equation (iii), 
a new Ernst potential can be written in the form
\begin{equation}
{\cal E}_S = e^{S^{(0)}}\frac{x(1+ab)+iy(b-a)-(1-ia)(1-ib)}
                         {x(1+ab)+iy(b-a)+(1-ia)(1-ib)},
\label{Ernst_GM}
\end{equation}
where $x$ and $y$ are the prolate-spheroidal coordinates:
$
\,\rho=\sigma\sqrt{x^2-1}\sqrt{1-y^2},\ z=\sigma xy\,,
$
with $\sigma>0$.
The ranges of $x$ and $y$ are $1 \leq x$ and $-1 \leq y \leq 1$.
The functions $a$ and $b$ satisfy the following 
simple first-order differential equations 
\begin{eqnarray}
\label{eq:ab}
(x-y)\del{x}a&=&
a\left[(xy-1)\del{x}S^{(0)}+(1-y^2)\del{y}S^{(0)}\right], \nonumber \\
(x-y)\del{y}a&=&
a\left[-(x^2-1)\del{x}S^{(0)}+(xy-1)\del{y}S^{(0)}\right], \nonumber \\
(x+y)\del{x}b&=&
-b\left[(xy+1)\del{x}S^{(0)}+(1-y^2)\del{y}S^{(0)}\right] , \nonumber\\
(x+y)\del{y}b&=&
-b\left[-(x^2-1)\del{x}S^{(0)}+(xy+1)\del{y}S^{(0)}\right]. \nonumber \\
\end{eqnarray}
The metric functions for the five-dimensional metric 
 (\ref{MBmetric}) are obtained
by using the formulas shown by \cite{ref7}, 
\begin{eqnarray}
e^{S}&=&e^{S^{(0)}}\frac{A}{B},   \label{e^S} \\
\omega&=&2\sigma e^{-S^{(0)}}\frac{C}{A}+C_1 , \label{omega}     \\
e^{2\gamma}&=&C_2(x^2-1)^{-1}A
                e^{2\gamma'}, \label{e_gamma}
\end{eqnarray}
where $C_1$ and $C_2$ are constants and
$A$, $B$ and $C$ are given by
\begin{eqnarray}
&&A:=(x^2-1)(1+ab)^2-(1-y^2)(b-a)^2 , \label{eq:A}\\
&&B:=[(x+1)+(x-1)ab]^2+[(1+y)a+(1-y)b]^2 \,, \nonumber \\ \label{eq:B}\\
&&C:=(x^2-1)(1+ab)[b-a-y(a+b)]  \nonumber
\\ &&\hskip 0.9cm
+(1-y^2)(b-a)[1+ab+x(1-ab)]\,.\label{eq:C}
\end{eqnarray}
The function $\gamma'$ in Eq. (\ref{e_gamma}) is a $\gamma$ function corresponding to the static metric,
\begin{eqnarray}
ds^2 &=& e^{-T^{(0)}}\left[
       -e^{2U^{\mbox{\tiny (BH)}}_0+S^{(0)}}(dx^0)^2
       +e^{-2U^{\mbox{\tiny (BH)}}_0-S^{(0)}}\rho^2(d\phi)^2  \right. \nonumber \\ && \left.
   +e^{2(\gamma'-U^{\mbox{\tiny (BH)}}_0)-S^{(0)}}\left(d\rho^2+dz^2\right) \right]
  +e^{2T^{(0)}}(d\psi)^2, \nonumber \\ \label{static_5}
\end{eqnarray}
where ${\displaystyle U_{0}^{\mbox{\tiny (BH)}}=\frac{1}{2}\ln\left( \frac{x-1}{x+1} \right)}$. 
And then the function $T$ is equals to $T^{(0)}$ and $U_1$ is given by 
the Einstein equation (vii).

Next we consider the solutions 
of the differential equations (\ref{eq:ab}). 
At first, we examine the case of a typical seed function 
\begin{equation}
S^{(0)}=\frac{1}{2}\ln\left[\,R_{d}+(z-d)\,\right], 
\label{seed_g}
\end{equation}
where  $R_{d}=\sqrt{\rho^2+(z-d)^2}$. 
The general seed function is composed of seed functions of this form.
Note that the above function $S^{(0)}$ is a Newtonian potential 
whose source is 
a semi-infinite thin rod \cite{ref8}.

In this case
we can confirm that the following $a$ and $b$ satisfy the 
differential equations (\ref{eq:ab}),
\begin{equation}
a=l_{\sigma}^{-1}e^{2\phi_{d,\sigma}}\ ,\ \ \ 
b=-l_{-\sigma}e^{-2\phi_{d,-\sigma}}\ ,\label{ab_phi}
\end{equation}
where
\begin{equation}
\phi_{d,c}=\frac{1}{2}
 \ln\left[\,e^{-\tilU_{d}}\left(e^{2U_c}+e^{2\tilU_{d}}\right)\,\right].
 \label{phi_i}
\end{equation}
Here the functions $\tilU_{d}$ and $U_{c}$ are defined as $\tilU_{d}:=\frac{1}{2}\ln\left[\,R_{d}+(z-d)\,\right]$ and 
$U_{c}:=\frac{1}{2}\ln\left[\,R_{c}-(z-c)\,\right]$.
Because of the linearity of the differential equations (\ref{eq:ab})
for $S^{(0)}$,
we can easily obtain $a$ and $b$ which correspond to a general seed function 
if it is a linear combination of (\ref{seed_g}).
See appendix \ref{app:NK} for this point.

The function $\gam'$ is defined from the static metric (\ref{static_5}), 
so that $\gam'$ obeys the following equations,
\begin{equation}
\del{\rho}\gam'=
\frac{1}{4}\rho\left[(\del{\rho}S')^2-(\del{z}S')^2\right]
 +\frac{3}{4}\rho\left[(\del{\rho}T')^2-(\del{z}T')^2\right],
 \label{eq:drho_gamma'}
\end{equation}
\begin{equation}
\del{z}\gam'=
\frac{1}{2}\rho\left[\del{\rho}S'\del{z}S'\right]
 +\frac{3}{2}\rho\left[\del{\rho}T'\del{z}T'\right],
 \label{eq:dz_gamma'}
\end{equation}
where the first terms are contributons from Eq. (iv) and the 
second terms come from Eq. (ii).
Here the functions $S'$ and $T'$ can be read out from Eq. (\ref{static_5}) as
\begin{eqnarray}
S'&=&2\,U^{(BH)}_0+S^{(0)},  \label{eq:S'} \\
T'&=&T^{(0)}. \label{eq:T'}
\end{eqnarray}
To integrate these equations we can use the following fact 
that, the partial differential equations
\begin{eqnarray}
\del{\rho}\gam'_{cd}
    &=&\rho\left[\del{\rho}\tU_{c}\del{\rho}\tU_{d}
            -\del{z}\tU_{c}\del{z}\tU_{d}\right],  \label{drho_gm}\\
\del{z}\gam'_{cd}
    &=&\rho\left[\del{\rho}\tU_{c}\del{z}\tU_{d}
           +\del{\rho}\tU_{d}\del{z}\tU_{c}\right], \label{dz_gm}
\end{eqnarray}
have the following solution, 
\begin{equation}
\gam'_{cd}=\frac{1}{2}\tU_{c}+\frac{1}{2}\tU_{d}-\frac{1}{4}\ln Y_{cd}, \label{gam'}
\end{equation}
where $Y_{cd}:=R_cR_d+(z-c)(z-d)+\rho^2$. The general solution of $\gamma'$
is given by the linear combination of the functions $\gamma'_{cd}$.


\section{application for simple seed metric}
\label{sec:application}

Using the solution-generating technique described above, we have obtained
a series of
five-dimensional axisymmetric stationary solutions in the previous paper
\cite{Mishima:2005id}.
In this section we retrace the analysis to derive the explicit form 
of the metric.

We adopt the five-dimensional 
Minkowski spacetime as a seed solution in this analysis. 
To obtain the solutions with sufficient variety, however,   
we add a freedom of one parameter to the seed metric and 
start from the following metric form, 
\begin{widetext}
\begin{eqnarray}
ds^2 &=&
\,-(dx^0)^2+\left(\sqrt{\rho^2+(z+\lam\sigma)^2}-(z+\lam\sigma) \right)d\phi^2
  +\left(\sqrt{\rho^2+(z+\lam\sigma)^2}+(z+\lam\sigma) \right)d\psi^2
   \nonumber \\
 && +\frac{1}{2\,\sqrt{\rho^2+(z+\lam\sigma)^2}}(d\rho^2+dz^2)
 \nonumber\\
 &=& -(dx^0)^2+\sigma\left(\sqrt{(x^2-1)(1-y^2)+(xy+\lambda)^2}
    -(xy+\lam) \right)d\phi^2
    +\sigma\Biglb(\sqrt{(x^2-1)(1-y^2)+(xy+\lam)^2}
   \nonumber \\&&
    +(xy+\lam) \Bigrb) d\psi^2
   +\frac{\sigma(x^2-y^2)}{2\,\sqrt{(x^2-1)(1-y^2)+(xy+\lam)^2}}
     \left[\frac{dx^2}{x^2-1}+\frac{dy^2}{1-y^2}\right],
 \label{eq:metirc_M}
\end{eqnarray}
\end{widetext}
where $\lam$ is an arbitrary real constant. To construct ringlike solutions we have to take $\lambda>1$.
By introducing the new coordinates $r$ and $\chi$: 
$$
\rho=r\,\chi,\ \ \ z=\frac{1}{2}(\chi^2-r^2)-\lam\sigma,
$$
we can easily confirm that the above metric corresponds to the Minkowski 
spacetime,
\begin{eqnarray}
ds^2 &=&
\,-(dx^0)^2+(dr^2+r^2d\phi^2)+(d\chi^2+\chi^2d\psi^2).
\end{eqnarray}
 From Eq.(\ref{eq:metirc_M}), 
we can read the form of the functions $S$ and $T$ 
as seed functions, 
\begin{eqnarray}
S^{(0)}=T^{(0)}&=&\tilU_{-\lambda \sigma} \nonumber \\ &=&
\frac{1}{2}\ln
      \left[\sqrt{\rho^2+(z+\lam\sigma)^2}+(z+\lam\sigma)\right] \nonumber \\
      &=& \frac{1}{2}\ln
      \Biglb[\sigma\left(\sqrt{(x^2-1)(1-y^2)+(xy+\lambda)^2}
       \right. \nonumber \\ && \left.
      +(xy+\lam) \right)\Bigrb] .
      \label{seed}
\end{eqnarray}
Using Eqs. (\ref{ab_phi}) and (\ref{phi_i}) 
we obtain the functions $a$ and $b$,  
\begin{eqnarray}
a
&=& \al\,\, \frac{(x-y+1+\lam)
         +\sqrt{x^2+y^2+2\lam xy+(\lam^2-1)}}
              {\ 2\left[(xy+\lam)+\sqrt{x^2+y^2+2\lam xy+(\lam^2-1)}
                                         \right]^{1/2}}, \nonumber \\
        \label{eq:a_pot}\\
b
&=& \beta\,\, \frac{\ 2\left[(xy+\lam)+\sqrt{x^2+y^2+2\lam xy+(\lam^2-1)}
                                         \right]^{1/2}}
                {(x+y-1+\lam)+\sqrt{x^2+y^2+2\lam xy+(\lam^2-1)}},\nonumber \\
    \label{eq:b_pot}
\end{eqnarray}
where $\al=2\sigma^{1/2} l^{-1}_{\sigma}$ 
and $\beta=-l_{-\sigma}/(2\sigma^{1/2})$. 

Next we reduce the explicit expression of the $\gamma'$.
When we substitute the seed functions (\ref{seed}) into Eqs. (\ref{eq:S'}) and (\ref{eq:T'}), 
the functions $S'$ and $T'$ are obtained as
\begin{eqnarray}
S'&=&2\,U^{(BH)}_0+S^{(0)}=2(\tilU_\sigma-\tilU_{-\sigma})
   +\tilU_{-\lam\sigma}   \\
T'&=&T^{(0)}=\tilU_{-\lam\sigma}.
\end{eqnarray}
As a result, the differential equations (\ref{eq:drho_gamma'}) and (\ref{eq:dz_gamma'})
become
\begin{eqnarray}
\del{\rho}\gam'
&=& \rho\left[(\del{\rho}\tilU_{\sigma})^2-(\del{z}\tilU_{\sigma})^2\right]
   +\rho\left[(\del{\rho}\tilU_{-\sigma})^2-(\del{z}\tilU_{-\sigma})^2\right]\nonumber \\
  &&
   +\rho\left[(\del{\rho}\tilU_{-\lam\sigma})^2-(\del{z}\tilU_{-\lam\sigma})^2
              \right]
      \nonumber \\
  && -2\rho\left[\del{\rho}\tilU_{\sigma}\del{\rho}\tilU_{-\sigma}
            -\del{z}\tilU_{\sigma}\del{z}\tilU_{-\sigma}\right] \nonumber \\
  &&
     +\rho\left[\del{\rho}\tilU_{\sigma}\del{\rho}\tilU_{-\lam\sigma}
            -\del{z}\tilU_{\sigma}\del{z}\tilU_{-\lam\sigma}\right]  \nonumber\\
  && -\rho\left[\del{\rho}\tilU_{-\sigma}\del{\rho}\tilU_{-\lam\sigma}
            -\del{z}\tilU_{-\sigma}\del{z}\tilU_{-\lam\sigma}\right] , \\
\del{z}\gam'
&=& 2\rho\left[\del{\rho}\tilU_{\sigma}\del{z}\tilU_{\sigma}\right]
   +2\rho\left[\del{\rho}\tilU_{-\sigma}\del{z}\tilU_{-\sigma}\right] \nonumber \\
  &&
   +2\rho\left[
              \del{\rho}\tilU_{-\lam\sigma}\del{z}\tilU_{-\lam\sigma}\right] 
                \nonumber \\
&&
-2\rho\left[\del{\rho}\tilU_{\sigma}\del{z}\tilU_{-\sigma}
           +\del{\rho}\tilU_{-\sigma}\del{z}\tilU_{\sigma}\right] \nonumber \\
  &&
+\rho\left[\del{\rho}\tilU_{\sigma}\del{z}\tilU_{-\lam\sigma}
           +\del{\rho}\tilU_{-\lam\sigma}\del{z}\tilU_{\sigma}\right] \nonumber \\
&&
-\rho\left[\del{\rho}\tilU_{-\sigma}\del{z}\tilU_{-\lam\sigma}
           +\del{\rho}\tilU_{-\lam\sigma}\del{z}\tilU_{-\sigma}\right] .
\end{eqnarray}
Now we divide $\gamma'$ into the six parts as
\begin{equation}
\gam' = \gam'_{\sigma,\sigma}+\gam'_{-\sigma,-\sigma}+\gam'_{-\lambda\sigma,-\lambda\sigma}-2\gam'_{\sigma,-\sigma}+\gam'_{\sigma,-\lambda\sigma}-\gam'_{-\sigma,-\lambda\sigma}  ,
\end{equation}
where the each terms are the solutions of Eqs. (\ref{drho_gm}) and (\ref{dz_gm}).
Finally, using the Eq. (\ref{gam'}), we obtain the resulting form of $\gamma'$ as
\begin{eqnarray}
\gam'
    &=& \frac{1}{2}\left( \tU_{\sigma}-\tU_{-\sigma} \right) 
\nonumber \\ &&
        +\tU_{-\lam\sigma}
        -\frac{1}{2}\left(
         \ln R_{\sigma}+\ln R_{-\sigma}+\ln R_{-\lam\sigma} 
         \right)  \nonumber  \\
    &&  +\frac{1}{4}\left(
                    2\ln Y_{\sigma,-\sigma}
                    -\ln Y_{\sigma,-\lam\sigma}
                    +\ln Y_{-\sigma,-\lam\sigma}
                    \right)-\frac{3}{4}\ln2. \nonumber \\
\end{eqnarray}

Consequently the functions which is needed to express the full metric
are completely obtained.
The full metric is expressed as 
\begin{eqnarray} 
ds^2 
&=&-\frac{A}{B}\left[dx^0-\left(2\sigma e^{-S^{(0)}}
         \frac{C}{A}+C_1\right) d\phi\right]^2    \nonumber  \\ &&
+\frac{B}{A}e^{-2S^{(0)}}\sigma^2(x^2-1)(1-y^2)
    (d\phi)^2  \nonumber \\ &&
+e^{2S^{(0)}}(d\psi)^2
+\frac{C_2\sigma}{\sqrt{2}(\lambda+1)} \nonumber \\ &&
\times
\frac{B(\sqrt{(x^2-1)(1-y^2)+(xy+\lambda)^2}+\lambda x +y)}
     {(x+1)\sqrt{(x^2-1)(1-y^2)+(xy+\lambda)^2}} \nonumber \\ &&
\times	
\left(\frac{dx^2}{x^2-1}+\frac{dy^2}{1-y^2}\right).
\label{full_metric}
\end{eqnarray}
Note that we have rewritten the second line of the metric (\ref{full_metric})
in the simpler form than the previous one.
In the following, 
the constants $C_1$ and $C_2$ are fixed as 
\[
C_1=\frac{\,\,2\sigma^{1/2}\,\al\,\,}{1+\al\beta},\ \ \ 
C_2=\frac{1}{\sqrt{2}(1+\al\beta)^2},
\]
to assure that the spacetime does not have global rotation
and that the periods of $\phi$ and $\psi$ 
become $2\pi$ at the infinity, respectively. 
The metric (\ref{full_metric}) in the canonical 
coordinates $\rho$ and $z$ is given in appendix \ref{app:canonical}.

\section{Properties of solution}
\label{sec:property}

\begin{figure}
  \vspace{0.5cm}
  \includegraphics[scale=0.28]{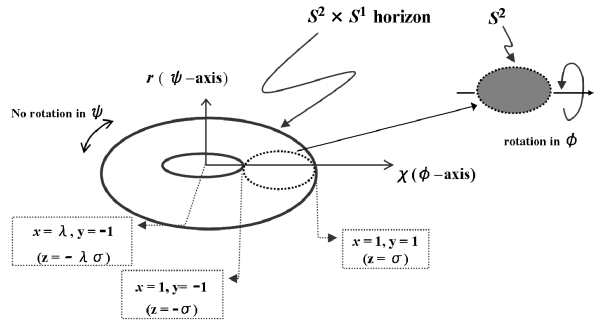}
  \caption{ 
   Schematic diagram of a local ringlike object which resides 
   in the spacetime. Generally some singular behavior appears 
   near the horizon.}
 \label{fig:ring}
 \end{figure}

In this section we investigate the qualitative features of 
the solution derived in the previous section. 
Before the detailed explanations, we summarize the basic properties 
of it briefly.
The spacetime described by this solution is axially symmetric, stationary,
and, in general, asymptotically flat.   
It is expected that there is a ringlike local object as in FIG. \ref{fig:ring}.
The parameters of $\lambda$ and $\sigma$ characterize the size and mass of the 
local object, respectively.  While appropriate combinations of
$\alpha$ and $\beta$ can be considered 
as the Kerr and the NUT parameters in four-dimensional case. 
Because of the existence of rotation, the spacetime has ergo-regions around
the ringlike object. Also, there 
are closed timelike curves in general case, where the metric component
$g_{\phi\phi}$ are negative.
In fact the value
of $g_{\phi\phi}$ can be negative around the inner disk of the ring at
$1<x<\lambda$ and $y=-1$ as we will show in the subsection \ref{sec:CTC}.
If we demand $g_{\phi\phi}=0$ at $1<x<\lambda$ and $y=-1$,
we obtain the following quadratic equation for $\beta$,
\begin{equation}
2\alpha \beta^2 + (2 + \alpha^2(\lambda+1))\beta +\alpha(\lambda-1)=0.
\label{eq:gpp0}
\end{equation}
We can confirm that the one of the solution of this equation 
\begin{equation}
\beta=-\frac{2+\al^2(\lam+1)-\sqrt{\al^4(\lam+1)^2-4\al^2(\lam-3)+4}}{4\al}
\label{noCTC}
\end{equation}
is the condition for $g_{\phi\phi}\ge 0$.
 Even in this case, there are conical singularities inside or outside the rings.
The reason for this is that the effect of rotation
cannot compensate for the gravitational attractive force.

In the rest of this section we 
discuss the physical properties of the solution in detail,
including the results given in the previous paper \cite{Mishima:2005id}.

\subsection{limits of solution}
The solution (\ref{full_metric}) has several limits which assist
our understanding of the nature of the solution.  
Direct generation of the limit solutions are given 
in appendix \ref{app:direct}.
One of the most considerable limits is
the Myers and Perry black hole with a single-rotation which is derived when
we set $\lambda=1$ and $\beta=0$. 
In fact the metric has the following expression, 
\begin{widetext}
\begin{eqnarray}
ds^2&=& -\frac{x-1-\alpha^2(1-y)}{x+1+\alpha^2(1+y)}
         \left(dx^0 + 2\sigma^{1/2}\alpha
          \frac{(1+\alpha^2)(1-y)}{x-1-\alpha^2(1-y)} d\phi\right)^2 
       +\sigma\frac{(x-1)(1-y)(x+1+\alpha^2(1+y))}{x-1-\alpha^2(1-y)} d\phi^2 
\nonumber \\  &&
       +\sigma(x+1)(1+y)d\psi^2
       +\frac{\sigma}{2}(x+1+\alpha^2(1+y))
         \left( \frac{dx^2}{x^2-1}+\frac{dy^2}{1-y^2}\right) \nonumber \\
 &=&-\frac{p^2x+q^2y-1}{p^2x+q^2y+1}
     \left( dx^0+2\sigma^{1/2}\, \frac{q}{p}\,\frac{1-y}{p^2x+q^2y-1}d\phi \right)^2  
       +\sigma\,\,\frac{p^2x+q^2y+1}{p^2x+q^2y-1}\,
        (x-1)\,(1-y)\, d\phi^2  %
 \nonumber   \\ &&
       +\sigma(x+1)\,(1+y)\,d\psi^2  
       +{\sigma}\frac{p^2x+q^2y+1}{2p^2}
     \left[ \frac{dx^2}{x^2-1}+\frac{dy^2}{1-y^2} \right],
     \label{eq:MPmetric_1}
\end{eqnarray}
\end{widetext}
where $p^2=1/(\al^2+1)$ and $q^2=\al^2/(\al^2+1)$.
Introduce new parameters $a_0$ and $m$, and new coordinates 
$\tilde{r}$ and $\theta$ 
through the relations, 
\begin{equation}
p^2=\frac{4\sigma}{m^2},
    \ \  q^2=\frac{a_0^2}{m^2},
\end{equation}
\begin{equation}
x=\frac{\tilde{r}^2}{2\sigma}-\lambda, \ y=\cos 2\theta,
\label{eq:rt_theta}
\end{equation}
so the metric (\ref{eq:MPmetric_1}) is transformed into
\begin{eqnarray}
ds^2 &=&-(1-\Delta)
    \left[dx^0+\frac{a_0\Delta\sin^2\theta}{1-\Delta}d\phi\right]^2 
    \nonumber \\
&& \hskip -0.3cm\,+\,\frac{1}{1-\Delta}
   \left[\tilde{r}^2+(m^2-a_0^2)\right]\sin^2\theta\, d\phi^2+\tilde{r}^2\cos^2\theta\,d\psi^2 
 \nonumber   \\
&& \hskip -0.3cm\,+\,(\tilde{r}^2+a_0^2\cos^2\theta)
        \left[d\theta^2+\frac{d\tilde{r}^2}{\tilde{r}^2-(m^2-a_0^2)}\right],
 \label{eq:MPmetric_2}
\end{eqnarray}
where $\Delta:=m^2/(\tilde{r}^2+a_0^2\cos^2\theta)$.
The line-element (\ref{eq:MPmetric_2}) is exactly the same form found 
by Myers and Perry.
 
Also this solution has a limit of a static black
ring or a rotational black string when the condition (\ref{noCTC}) holds. 
The former case is realized when 
we take the limit $\al \rightarrow 0$.
The parameter $\beta$ also approaches $0$ in this limit.
Therefore the function $C$ and the constant $C_1$ become $0$ and then
the spacetime approaches static one in this limit. 
In addition, the functions $A$ and $B$
become simple forms
\begin{equation}
 A = x^2 -1, ~~~B=(x+1)^2.
\end{equation}
So, the ergo-region and the CTC region do not appear in this spacetime obviously.
The metric form of this limit becomes
\begin{widetext}
\begin{eqnarray}
 ds^2 &=& - \frac{x-1}{x+1} (dx^0)^2 
       + \sigma \frac{(x+1)^2 (1-y^2)}{\sqrt{(x^2-1)(1-y^2) 
                       + (xy + \lambda)^2}+ (xy + \lambda)} d\phi^2 
      \nonumber \\
      && + \sigma(\sqrt{(x^2-1)(1-y^2)+ (xy + \lambda)^2}- 
                    (xy + \lambda)) d\psi^2
      \nonumber \\
      && + \frac{\sigma}{2(\lambda + 1)}
         \frac{(x+1)(\sqrt{(x^2-1)(1-y^2)+ (xy + \lambda)^2}
            +\lambda x + y))}{\sqrt{(x^2-1)(1-y^2)+ (xy + \lambda)^2}}
        \left(\frac{dx^2}{x^2-1}+\frac{dy^2}{1-y^2}\right).
\end{eqnarray}
\end{widetext}
This metric can be written in canonical coordinates as Eq. (\ref{eq:static_cano}) in Appendix \ref{app:canonical}.

The latter is realized
when the parameter $\lambda$ goes to infinity
under the condition: $\al = \tilde{\al}\times\sqrt{2/\lambda}$
with $-1<\tilde{\al}<1$. In this case $\beta$ goes to infinity like $-\tilde{\alpha}\times \sqrt{\lambda/2}$ while
the product $\alpha\beta$ is finite and $-1<\alpha\beta=-\tilde{\alpha}^2<0$.
As a result, the functions $a$ and $b$ approaches constants $\tilde{\alpha}$ and
$-\tilde{\alpha}$, respectively. 
The $\psi$-$\psi$ component of the metric diverges 
except at $x=\infty$ and $y=-1$.
To avoid this singular behavior we have to replace the angular coordinate $\psi$
with $\frac{\tilde{\psi}}{\sqrt{2 \lambda}}$.  Also
we have to rescale $\phi$ as $\phi=\sqrt{2\lambda}\tilde{\phi}$.
After these replacements, the metric can be rewtitten as
\begin{widetext}
\begin{eqnarray}
 ds^2 &=&-\frac{(1-\at^2)^2(x^2-1)-4\at^2(1-y^2)}{(x+1-\at^2 (x-1))^2 
         + 4\at^2 y^2}
         \left(dx^0  
          + 2\sigma^{1/2} 
           \frac{2 \at(1+\at^2)((1-\at^2)x+1+\at^2)(1-y^2)}
                {(1-\at^2)^2((1- \at^2)(x^2-1)-4\at^2 (1-y^2))} 
                                       d\tilde{\phi}\right)^2 \nonumber \\
      && +\sigma \frac{(x^2-1)(1-y^2)(x+1-\at^2 (x-1))^2 + 4\at^2 y^2}
                      {(1- \at^2)^2(x^2-1)-4\at^2 (1-y^2)}  d\tilde{\phi}^2 
         +\sigma d\tilde{\psi}^2 
 \nonumber \\ && 
         +\sigma \frac{(x+1-\at^2(x-1))^2+4\at^2y^2}{2(1-\at^2)^2}
         \left[ \frac{dx^2}{x^2-1}+\frac{dy^2}{1-y^2} \right],  \nonumber \\
     &=& -\frac{\pt^2x^2+\qt^2y^2-1}{(\pt x+1)^2 + \qt^2 y^2}
            \left( dx^0 
    +2\sigma^{1/2}\frac{2\at(\pt x+1)(1- y^2)}{\pt(\pt^2x^2+\qt^2y^2-1)}
               d\tilde{\phi}\right)^2  \nonumber \\
      && +\sigma \frac{(x^2-1)(1-y^2)(\pt x+1)^2 + \qt^2 y^2}{\pt^2x^2+\qt^2y^2-1}
          d\tilde{\phi}^2  +\sigma d\tilde{\psi}^2 
        +\sigma \frac{(\pt x+1)^2 + \qt^2 y^2}{2\pt^2}
       \left[ \frac{dx^2}{x^2-1}+\frac{dy^2}{1-y^2} \right],
       \label{eq:metric_BS}
\end{eqnarray}
\end{widetext}
where $\pt=\frac{1-\at^2}{1+\at^2}$ and $\qt=\frac{2\at^2}{1+\at^2}$.
This is just a rotational black string metric.

\subsection{asymptotic flatness}
It should be noted that
the solution-generating techniques developed in this study have an advantage that the resulting five-dimensional solutions hold asymptotic
flatness
if we adopt a five-dimensional asymptotically flat seed solution.
This can be easily confirmed by the rod structure analysis, 
which will be discussed in the next subsection.
If we take the asymptotic limit, $x \rightarrow \infty$,
in the prolate-spheroidal coordinates, the 
metric form (\ref{full_metric}) approaches the asymptotic form of 
the Minkowski metric (\ref{eq:metirc_M}),
\begin{eqnarray}
 ds^2 &\sim& -(dx^0)^2+ \sigma x(1-y)d\phi^2 +\sigma x(1+y)d\psi^2 
\nonumber \\ &&
 +\frac{\sigma}{2x}{dx^2}
 +\frac{\sigma x}{2(1-y^2)}{dy^2}.
\end{eqnarray}
Also the asymptotic form of $\mE_{S}$ near the infinity $x=\infty$ becomes 
\begin{eqnarray}
\mE_{S}&=&\tilde{r}\cos\theta\,
\left[\,1\,-\,\frac{\sigma}{\tilde{r}^2}\,\frac{P(\al,\beta,\lam)}
                                       {(1+\alpha\beta)^2}
     \,+\cdots\right] \nonumber  \\ && 
     +2\,i\,\sigma^{1/2}\,\left[\,\frac{\alpha}{1+\alpha\beta}
      \,-\,\frac{2\sigma\cos^2\theta}{\tilde{r}^2}\,\frac{Q(\al,\beta,\lam)}
                                                 {(1+\alpha\beta)^3}
    \,\,+\cdots\,\right], \nonumber \\
\end{eqnarray}
where 
\begin{eqnarray*}
P(\al,\beta,\lam)
   &=&  4(1 + \alpha^2 - \alpha^2 \beta^2)       \\
Q(\al,\beta,\lam)
   &=& \alpha(2\alpha^2-\lam+3)-2\alpha^2\beta^3  \\
    &&\hskip -0.4cm
  -\beta\left[2(2\alpha\beta+1)(\alpha^2+1)+(\lam-1)\al^2(\al\beta+2)\right],
\end{eqnarray*}
and we use the coordinates $(\tilde{r},\theta)$ through Eq. (\ref{eq:rt_theta}).
This fact means that, even if $\lam\neq 1$ or $\beta\neq 0$, 
the asymptotic form has the same asymptotic behavior as the case with $\lam=1$ and $\beta=0$, i.e., the Myers-Perry black hole. 
 From the asymptotic behavior, we can compute
the mass parameter $m^2$ and rotational parameter $m^2a_0$:
\begin{equation}
m^2=\sigma \frac{P(\al,\beta,\lam)}{(1+\alpha\beta)^2},\ \ 
m^2a_0=4\sigma^{3/2}\frac{Q(\al,\beta,\lam)}{(1+\alpha\beta)^3}.
\label{mass}
\end{equation}

\subsection{rod structure analysis}

We analyze the rod structure of the solution,
which was studied for the higher-dimensional Weyl solutions
by  Emparan and Reall \cite{ref8} and for the nonstatic solutions
by Harmark \cite{refHAR}. 
The brief review of these
methods are given in appendix \ref{app:rod}.
We have four rods whose intervals are
$z\in[-\infty,-\lambda \sigma]$, $[-\lambda \sigma,-\sigma]$, 
$[-\sigma,\sigma]$, and $[\sigma,\infty]$ at $\rho=0$
which correspond with $(x,y) \in \{\lambda \le x, y=-1\}$, 
$\{1 \le x \le \lambda, y=-1\}$,
$\{x=1, -1\le y \le 1\}$, and $\{1 \le x, y=1\}$,
 respectively.
The semi-infinite rod $[-\infty,-\lambda \sigma]$ has the 
direction ${\bf v}=(0,0,1)$. Therefore this rod corresponds to the fixed points of the $\psi$-rotation.
The finite rod $[-\lambda \sigma,-\sigma]$ has the direction
\begin{eqnarray}
{\bf v} &=& (\Xi,1,0), \nonumber \\ 
\Xi  &=& \frac{2\sqrt{\sigma}(2\alpha \beta^2 + (2 + \alpha^2(\lambda+1))\beta +\alpha(\lambda-1))}{(1+\alpha\beta)(\lambda-1+\alpha\beta(\lambda+1))}. \nonumber \\
\label{eq:direction_phi}
\end{eqnarray}
It can be shown that this rod is spacelike.
In general it does not correspond to the fixed points of the $\phi$-rotation.
When the condition (\ref{eq:gpp0}) holds, 
it becomes the fixed points of the $\phi$-rotation.
The finite rod $[-\sigma,\sigma]$
has the direction
\begin{equation}
{\bf v} = (1,\Omega,0), ~~~ 
\Omega =\frac{(1+\alpha\beta)((\lambda+1)\alpha-2\beta)}
             {2\sqrt{\sigma}((\lambda+1)\alpha^2 + 2)},
\label{eq:direction_t}
\end{equation}
which corresponds to the region of time translational invariance.
The semi-infinite rod $[\sigma,\infty]$  has the 
direction ${\bf v}=(0,1,0)$. Therefore this rod corresponds to the fixed points of the $\phi$-rotation. When the condition (\ref{eq:gpp0}) holds,
the topology of the event horizon is $S^1 \times S^2$ 
for $\lambda>1$ as in Figure 1 because the rod $[-\sigma,\sigma]$ has
the rods in the $\partial/\partial \phi$ direction on  each side.
Also the solution is free of the pathology of the Dirac-Misner string \cite{Elvang:2004xi} in this case.

\begin{figure}
  \includegraphics[scale=0.17,angle=0]{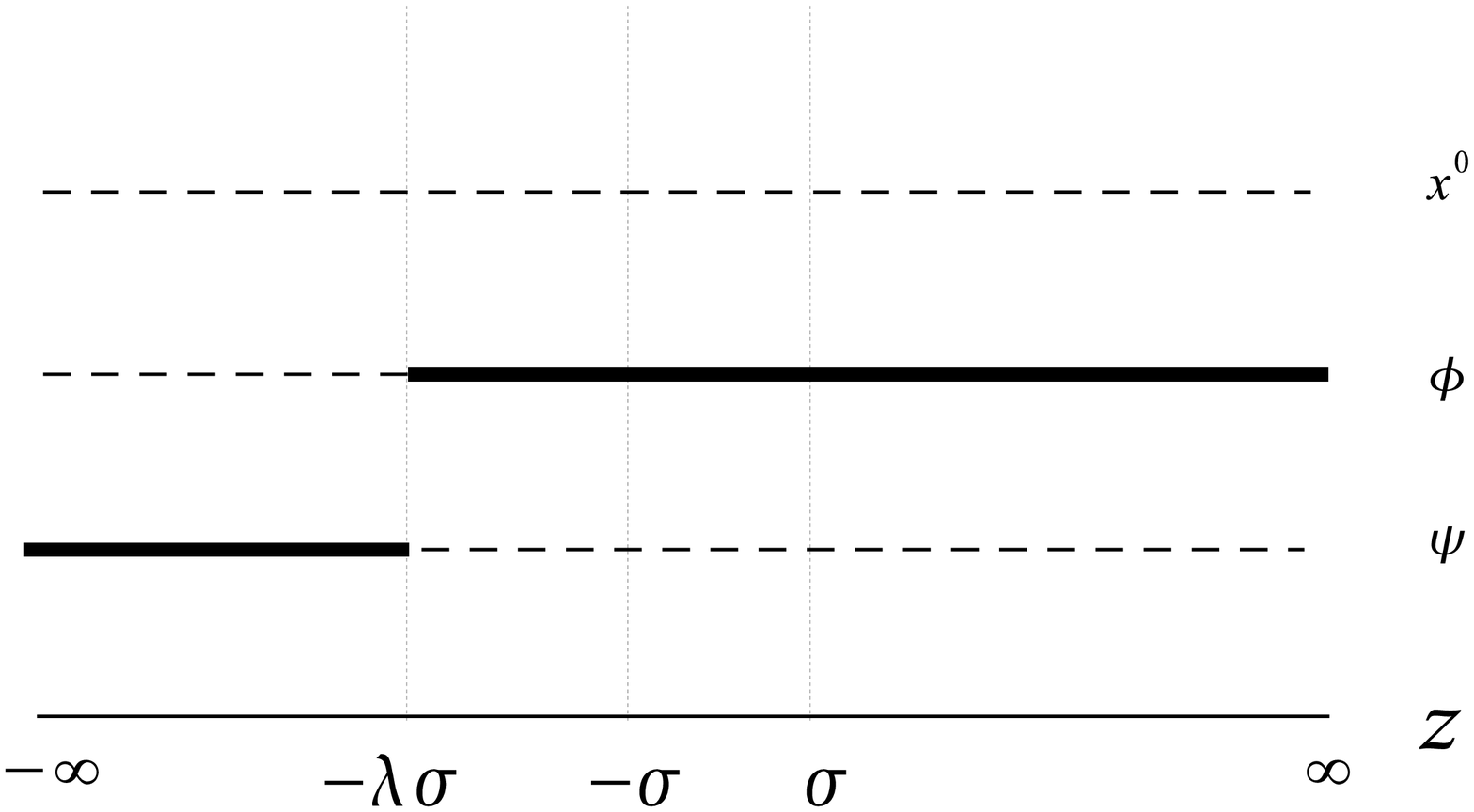}
  \includegraphics[scale=0.17,angle=0]{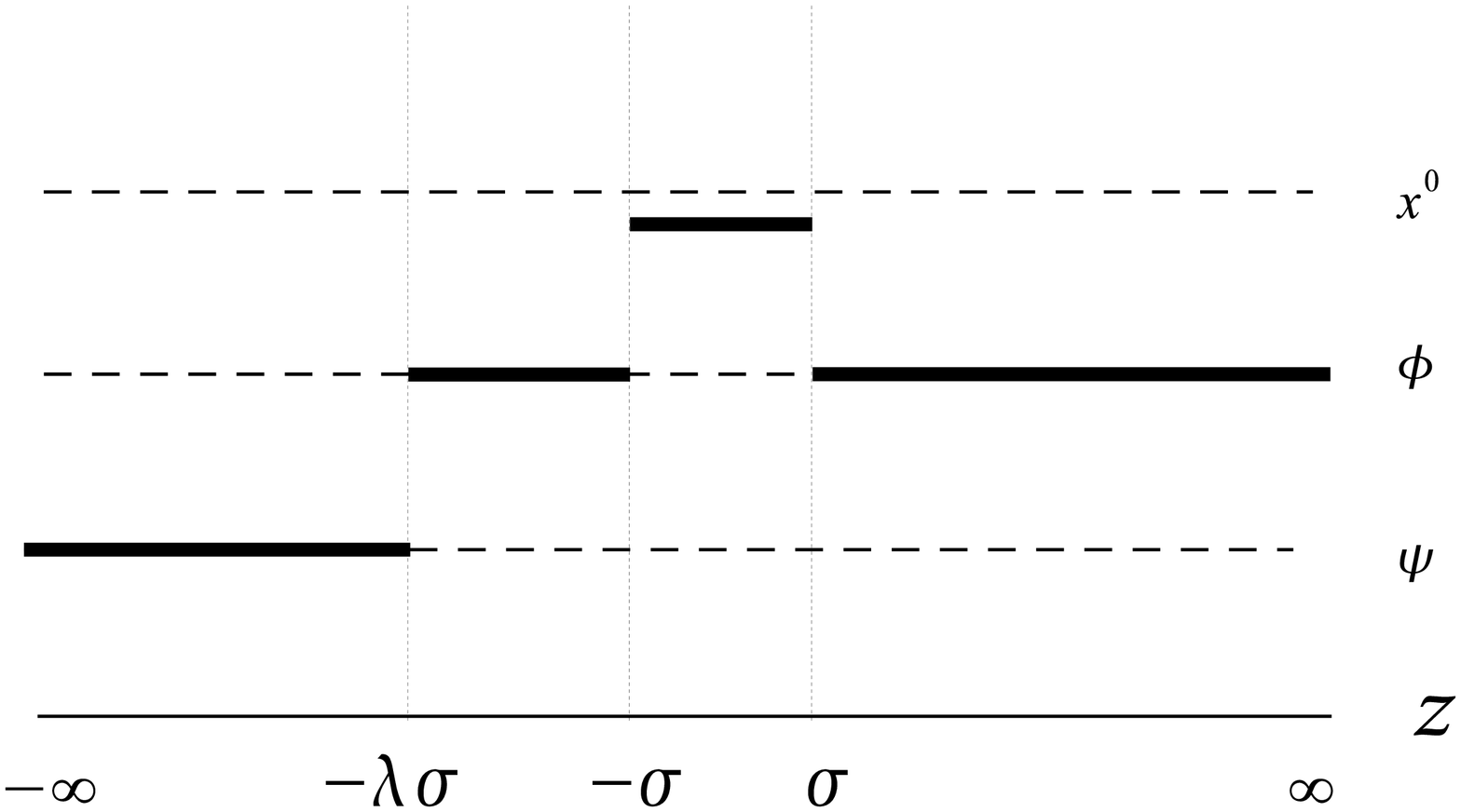}
  \caption{Schematic pictures of rod structures. The left panel
shows the rod struture of Minkowski spacetime which is a seed of 
$S^2$-rotating black ring. The right panel shows the rod structure
of the $S^2$-rotating black ring. The segment $[-\sigma,\sigma]$
of semi-infinite rod of the seed 
becomes finite timelike rod and changes its direction
 by the solution-generating transformation.
Here we put the finite timelike rod between $x^0$ and $\phi$ lines
because the corresponding eigenvector has nonzero $x^0$ and $\phi$ components.
Note that the segment $[-\lambda \sigma,-\sigma]$ also
changes its direction by the transformation in general.}
 \label{fig:rods_S2}
 \end{figure}

We show the schematic pictures of rod structures of 
the $S^2$-rotating black ring and its seed solution 
in Fig. \ref{fig:rods_S2}.
By the solution-generating transformation
the segment $[-\sigma,\sigma]$
of semi-infinite spacelike rod of the seed,
which corresponds to the fixed point of $\phi$-rotation,
turns into the finite timelike rod with the direction
(\ref{eq:direction_t}). To indicate that this vector has nonzero
$x^0$ and $\phi$ components, the rod is laid between $x^0$ and $\phi$
axes in Fig. \ref{fig:rods_S2}.
In general the segment $[-\lambda \sigma,-\sigma]$ also
changes its direction from $\partial/\partial \phi$ to
(\ref{eq:direction_phi}).
We can see that
the solitonic transformation keeps the existence of the two semi-infinite
spacelike rods intact. This fact assures the asymptotic flatness of the obtained solution.

\subsection{ergo region}
\begin{figure}
  \vspace{0.5cm}
  \includegraphics[scale=0.7]{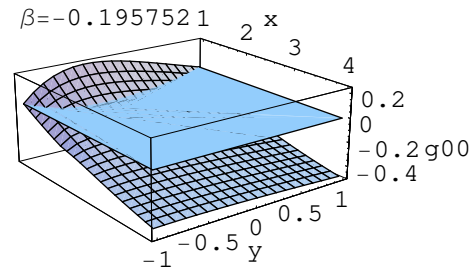}
  \caption{ 
   The behavior of $0$-$0$ component of the metric (\ref{full_metric})
   in the case of $(\alpha,\, \beta,\, \lambda)=(0.5,\, -0.195752,\,2)$ .
   The region where the component function is above the level zero 
   corresponds to the ergo-region.}
 \label{fig:ergo}   
 \end{figure}

As naturally expected from the presence of the rotation, the new solutions
 have ergo-regions where $g_{00} > 0 $. 
In fact, the 0-0 component of the metric (\ref{full_metric})
becomes positive near $x=1$ because the function $A$ becomes
negative there. The form of this componet
at $x=1$ is obtained as
\begin{equation}
 g_{00} = \frac{((\lambda+1)\alpha-2\beta)^2(1-y^2)}
{8(\lambda+y)+(2\beta(1-y)+\alpha(\lambda+1)(1+y))^2}.
\end{equation}
Figure \ref{fig:ergo} is the plot of  $g_{00}$
for the region $1<x<4$ and $-1<y<1$ in a typical case of $(\alpha,\, \beta,\, \lambda)=(0.5,\, -0.195752,\,2)$ which satisfies the condition (\ref{noCTC}). 
There exists an ergo-region around the event horizon $x=1$.
 
\begin{figure*}
  \includegraphics[scale=0.4]{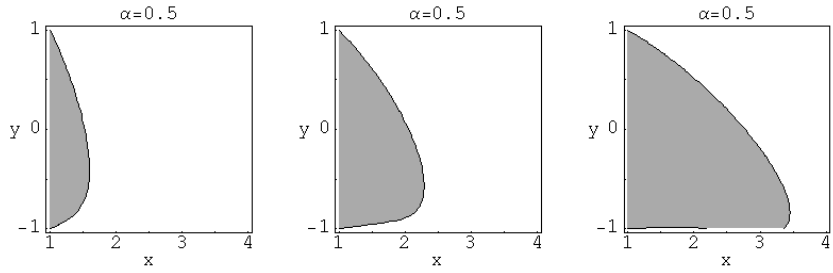}
  \caption{ 
   Ergo-regions for the cases of $\alpha=0.5,0.7,0.9$ with $\lambda=2$.
   The values of $\beta$ are determined by Eq. (\ref{noCTC}).
   In the shaded regions the values of $g_{00}$ are positive.}
 \label{fig:ergo2}   
 \end{figure*}

Next we consider the relations between the ergo-region and the parameter $\alpha$.
We plot the ergo-regions for the cases of $\alpha = 0.5, 0.7, 0.9$ 
with $\lambda = 2$ in FIG. \ref{fig:ergo2}. The values of $\beta$ are determined by the condition
 (\ref{noCTC}). 
The ergo-region of this ring spreads out towards the nonrotational axis 
$(x \ge \lambda,y=-1)$ of 
the ring as the value of $\alpha$ becomes large.

 \begin{figure*}
  \includegraphics[scale=0.5]{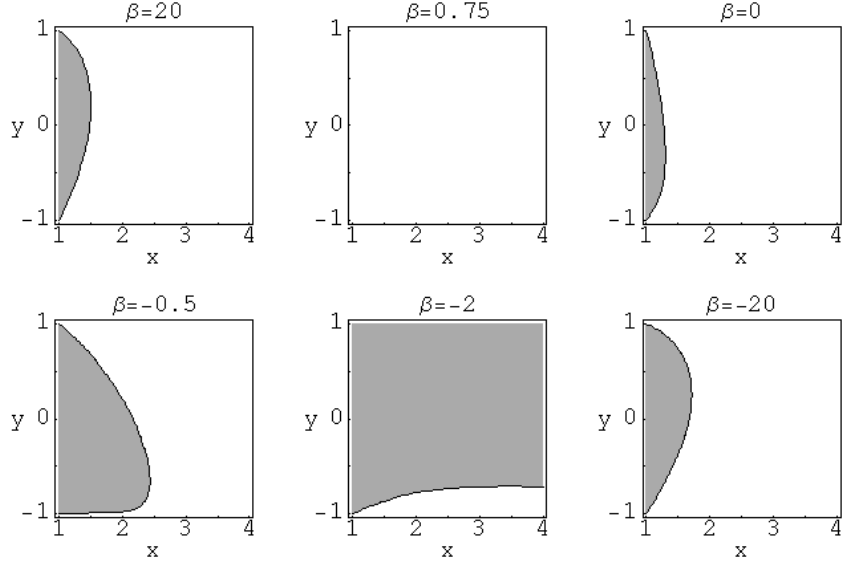}
  \caption{ 
   Ergo-regions for singular cases where $\alpha=0.5$ and
   $\lambda=2$. In the shaded regions the values of $g_{00}$ become positive.
   The values of $\beta$ do not satisfy  Eq. (\ref{noCTC}). 
   When $\beta=\frac{\alpha(1+\lambda)}{2}$, the ergo-region disappears.
   }
 \label{fig:ergo3}   
 \end{figure*}

For singular cases where Eq. (\ref{noCTC}) does not hold, we investigate 
the behaviors of $g_{00}$ for different values of $\beta$ as in FIG. 
\ref{fig:ergo3}.  For the cases of large absolute values of $\beta$,
the shapes of ergo-regions are similar with each other. There are two special cases,
where $\beta=\frac{\alpha(1+\lambda)}{2}$ and $\beta=-\frac{1}{\alpha}$.
The former case does not have ergo-regions.
The latter case would be singular because the mass and rotational parameters 
diverges.

\subsection{closed timelike curve}
\label{sec:CTC}

\begin{figure*}
  \includegraphics[scale=0.8]{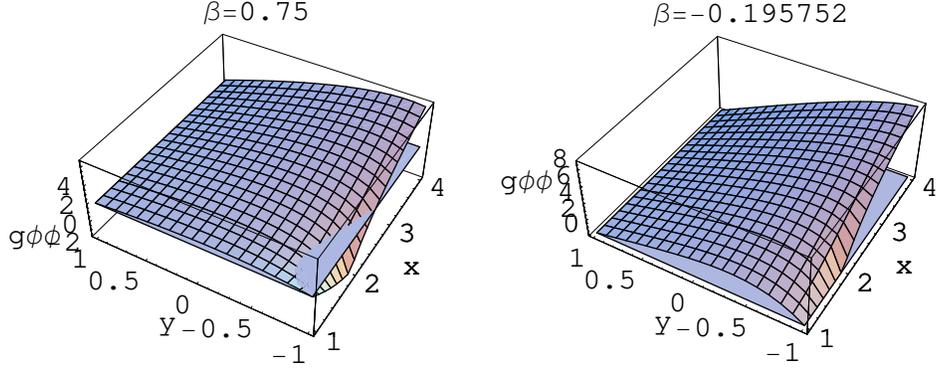}
  \caption{ 
   The behavior of $\phi$-$\phi$ component of the metric 
   in the cases of $(\alpha,\, \beta,\, \lambda)=(0.5,0.75,2)$ and
   $(0.5, -0.195752,2)$ which satisfies the Eq. (\ref{noCTC}).
   The corresponding component in the right panel is always non-negative, 
   while for the general case the component becomes negative near the horizon, 
   which means the existence of CTC-regions.}
   \label{fig:gpp}
 \end{figure*}

There may exist closed timelike curves in this spacetime. It would be exist 
if the metric function $g_{\phi\phi}$ becomes negative.
At first it can be easily
shown that the value of $g_{\phi\phi}$ is zero at $y=1$. There is no harmful feature around there.
However we can confirm the appearance of CTC from the fact that
this component becomes
\begin{widetext}
\begin{equation}
 g_{\phi\phi}=-\frac{4\sigma(2\alpha \beta^2 + 
        (2 + \alpha^2(\lambda+1))\beta +\alpha(\lambda-1))^2(x^2-1)}
       {(1+\alpha\beta)^2(8\beta^2(\lambda-x)
          +((\lambda-1)(x+1)+ \alpha\beta(\lambda+1)(x-1))^2)},
\label{eq:gpp1}
\end{equation}
for the range $1<x<\lambda$ at $y=-1$.
This value is always negative except when the parameters satisfy 
the condition (\ref{eq:gpp0}).
When $\lambda$ and $\alpha$ are given, the parameter $\beta$ must be
\begin{equation}
\beta=\beta_{+}
     =-\frac{2+\al^2(\lam+1)+\sqrt{\al^4(\lam+1)^2-4\al^2(\lam-3)+4}}{4\al},
\label{eq:conj}
\end{equation}
or
\begin{equation}
\beta=\beta_{-}
     =-\frac{2+\al^2(\lam+1)-\sqrt{\al^4(\lam+1)^2-4\al^2(\lam-3)+4}}{4\al},
\label{eq:noCTC}
\end{equation}
\end{widetext}
to satisfies the condition (\ref{eq:gpp0}).
Even in this case there can appear the CTC 
when the function $B$ becomes sufficiently small outside the
ergo-region.
We can show that the value of $B$ becomes zero at
\begin{equation}
 x=\frac{(\lambda^2-1)\alpha^2-4\beta^2}{4 \alpha\beta}, ~~~ y=0.
\label{eq:zero_point}
\end{equation}
For $\beta=\beta_{+}$, the coordinate value $x$ of (\ref{eq:zero_point}) 
is in the range $x>1$. 
Therefore there appears singular behavior and $g_{\phi\phi}$ becomes
negative in its neighborhood.
While, when $\beta=\beta_{-}$, this singular behavior does not
appear because the value of $x$ in (\ref{eq:zero_point}) is less than 1.
As a result, the condition (\ref{eq:noCTC})
makes the singular structure of the spacetimes fairly mild
as seen in the right panel of FIG. \ref{fig:gpp}. 
The general case has regions where $g_{\phi\phi}$ becomes negative 
as in the left panel of  FIG. \ref{fig:gpp}.


\begin{figure*}
  \includegraphics[scale=0.55]{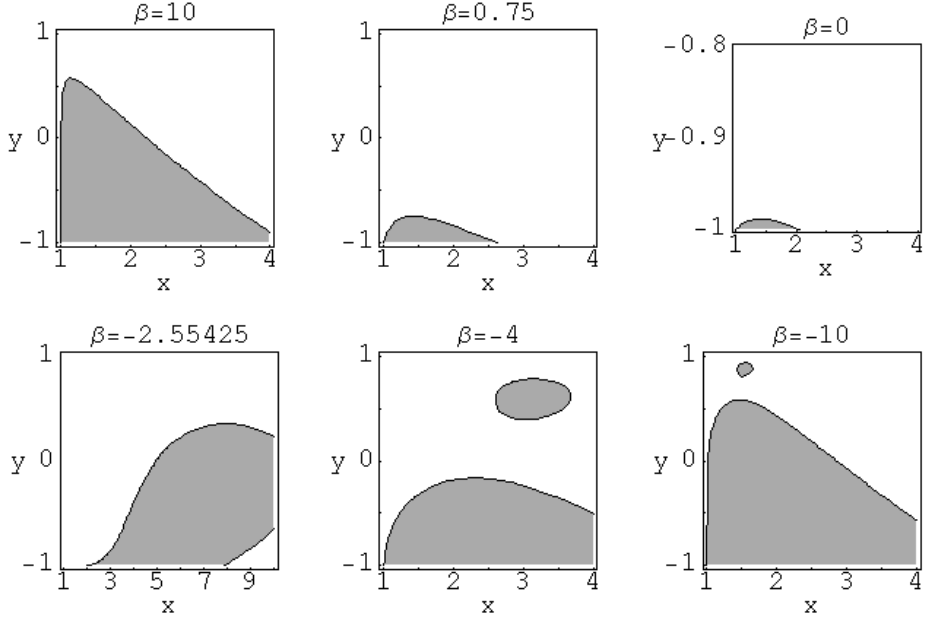}  
  \caption{ 
  CTC-regions. In the shaded regions
  the values of $g_{\phi\phi}$ are negative. Note that the $y$-range of
  the right-upper panel and the $x$-range of the left-lower panel 
 are different from the others. The left-lower panel corresponds to 
 $\beta=\beta_{+}$.
   }
   \label{fig:gpp2}
 \end{figure*}

We show  
the regions where $g_{\phi\phi}<0$ for different values of $\beta$ 
in FIG. \ref{fig:gpp2}. 
Note that the CTC-region can not touch with the event horizon $x=1$
except for the inner edge of the ring, $(x,y)=(1,-1)$. 
When the absolute values of $\beta$  are large,
these regions become similar shapes with each other.

\subsection{excess (deficit) angles}
Even if the closed timelike curve does not exist, i.e., $\beta=\beta_{-}$,
there exists a kind of strut structure in this spacetime.
The reason for this is that the effect of rotation
cannot compensate for the gravitational attractive force.
The periods of the coordinates $\phi$ and $\psi$ should be defined as 
\begin{equation}
 \Delta \phi = 2 \pi \lim_{\rho \rightarrow 0} \sqrt{\frac{\rho^2 g_{\rho\rho}}{g_{\phi\phi}}} 
 ~~~\mbox{and}~~~
 \Delta \psi = 2 \pi \lim_{\rho \rightarrow 0} \sqrt{\frac{\rho^2 g_{\rho\rho}}{g_{\psi\psi}}}
\end{equation}
to avoid a conical singularity.
Both the value of $\Delta \psi$ for $x>\lambda$ and $y=-1$ and the value of
$\Delta \phi$ for $y=1$, i.e., outside part of the $\phi$-axis-plane, are $2\pi$. 
While the period of $\phi$ inside the ring 
can be defined only when the condition (\ref{eq:gpp0}) holds.
In this case the period becomes
\begin{equation}
 \Delta \phi = 2 \pi \frac{\lambda -1 
 +(\lambda + 1)\alpha \beta}{\sqrt{\lambda^2-1}(1+\alpha\beta)}, \label{in_phi}
\end{equation}
which is less than $2\pi$ for $\beta=\beta_{-}$
and larger than $2\pi$ for $\beta=\beta_{+}$.
Hence, two-dimensional disklike struts, which appear in the case of 
static black rings \cite{ref8}, are needed to prevent the collapse of the 
$S^2$-rotating black rings. 


We have introduced four parameters $\lambda$, $\sigma$, $\alpha$ and $\beta$ 
in our analysis. Also, we need the condition (\ref{noCTC}) for 
the disappearance of 
CTC regions. As a result there are three independent parameters
for the $S^2$-rotating black ring. 
Here we take $\lambda$, $m$ and 
\begin{equation}
h:=\frac{\lambda -1 
 +(\lambda + 1)\alpha \beta}{\sqrt{\lambda^2-1}(1+\alpha\beta)} \label{def:h}
\end{equation}
as these physical parameters. From Eqs. (\ref{eq:gpp0}), (\ref{mass})
and (\ref{def:h}) we can obtain the relations between these parameters and 
the other parameters $\sigma$, $\alpha$ and $\beta$ as
\begin{eqnarray}
 \sigma &=& \frac{m^2 h}{2\sqrt{\lambda^2 -1}(1-h^2)}, \\
 \alpha &=& \pm \sqrt{\frac{2}{h(\lambda+1)}
    \frac{\sqrt{\lambda-1}-h\sqrt{\lambda+1}}
    {\sqrt{\lambda+1}-h\sqrt{\lambda-1}}},\\
 \beta &=& \mp \sqrt{\frac{h(\lambda-1)}{2}
   \frac{\sqrt{\lambda-1}-h\sqrt{\lambda+1}}
   {\sqrt{\lambda+1}-h\sqrt{\lambda-1}}}.   
\end{eqnarray}
The condition that the parameters $\alpha$ and $\beta$ should be real is
\begin{equation}
 0<h<\sqrt{\frac{\lambda-1}{\lambda+1}}<1
\end{equation}
or
\begin{equation}
 h>\sqrt{\frac{\lambda+1}{\lambda-1}}>1.
\end{equation}
The former case corresponds to the case of $\beta=\beta_-$
and the latter to $\beta=\beta_+$.
Therefore the period of $\phi$ inside the ring 
is always larger than $2\pi$ when the condition $\beta=\beta_+$ holds
which corespond to the left-lower panel of FIG. \ref{fig:gpp2}.
It should be noted that the mass parameter $m^2$ 
is negative in this case.
%

\subsection{maximal rotation limit} 
In this subsection we investigate the rotational parameter $a_0$
for the $S^2$-rotating black ring.
The rotational parameter $a_0$ in Eq. (\ref{mass}) 
can be rewritten by using the parameters $\lambda$, $m$ and $h$ as
\begin{equation}
 \frac{a_0^2}{m^2}= \frac{1-h}{1+h}
              -\frac{2h}{1-h^2}\left(\sqrt{\frac{\lambda^2}{\lambda^2-1}}-1\right).
\end{equation}
When we fix the parameters $m$ and $h$, the rotational parameter
increases uniformly according to the value of $\lambda$ and has a
maximum value
\begin{equation}
 {a_0}_{max} 
             = \sqrt{\frac{m^2(1- h)}{1+h}}.
\end{equation}
The parameter $\lambda$ diverges at the maximum of $a_0$,  
while 
$\sigma$ goes to $0$ with keeping the value of $\lambda \sigma$ finite. 
Then the physical size of the ring is kept finite.
The parameter $\alpha$ goes to $0$  at the maximum of $a_0$, while 
$\beta$ diverges to infinity. 
The parameters behave around the maximum of $a_0$ as
\begin{eqnarray}
 &&\sigma \sim \frac{m^2 h}{2\lambda(1-h^2)}, ~~~~~
 \alpha \sim \pm \sqrt{\frac{2}{h\lambda}}, \nonumber \\ &&
 \beta \sim \mp \sqrt{\frac{h\lambda}{2}}, ~~~~~
 1+\alpha\beta \sim \frac{2}{\lambda (1-h)} .
 \label{eq:para_extreme}
\end{eqnarray}
When we take this limit, we have to redefine the coordinate $x$ as,
for example, $\tilde{x}-1 = (x-1)/\lambda$ to extract the regular 
form of the solution.

\begin{figure}
  \includegraphics[scale=0.28]{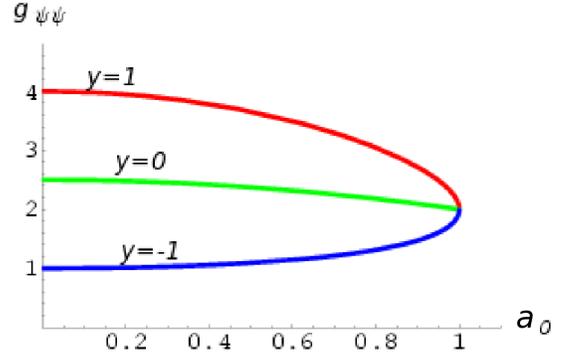}
  \caption{ 
  Plots of $\psi$-$\psi$ components of the $S^2$-rotating black ring for 
  $a_0$ with $m^2=3$ and $h=0.5$. 
  Upper, middle, and lower lines correspond with
  outer edge $(y=1)$, middle $(y=0)$
   and inner edge $(y=-1)$ of the ring, respectively. 
   } \label{lambda}
 \end{figure}

In FIG. \ref{lambda}, we plot the values of $g_{\psi\psi}$
of the event horizon at $y=-1$ (inner edge), 
$y=0$ (middle) and $y=1$ (outer edge). Here we set the parameters as  $m^2=3$ and $h=0.5$. 
The circumferences of the inner and the outer edge of the ring approach
each other as the parameter $a_0$ becomes large. 
As we will see in \ref{sec:C-metric}, the event horizon 
degenerates at the maximum of $a_0$.
Then we call this limit
the extreme limit of the solution. 
In fact the rotational parameter ${a_0}_{max}$ equals the mass parameter
$m$ when we take the five-dimensional Kerr black hole limit $h \rightarrow 0$.


\subsection{curvature invariants}
In this section we consider the possibility of the 
naked curvature singularity.
Here we examine the scalar curvature 
\begin{eqnarray}
 K &=& R_{ijkl}R^{ijkl},
\end{eqnarray}
which is usually called Kretchman invariant.

\begin{figure*}
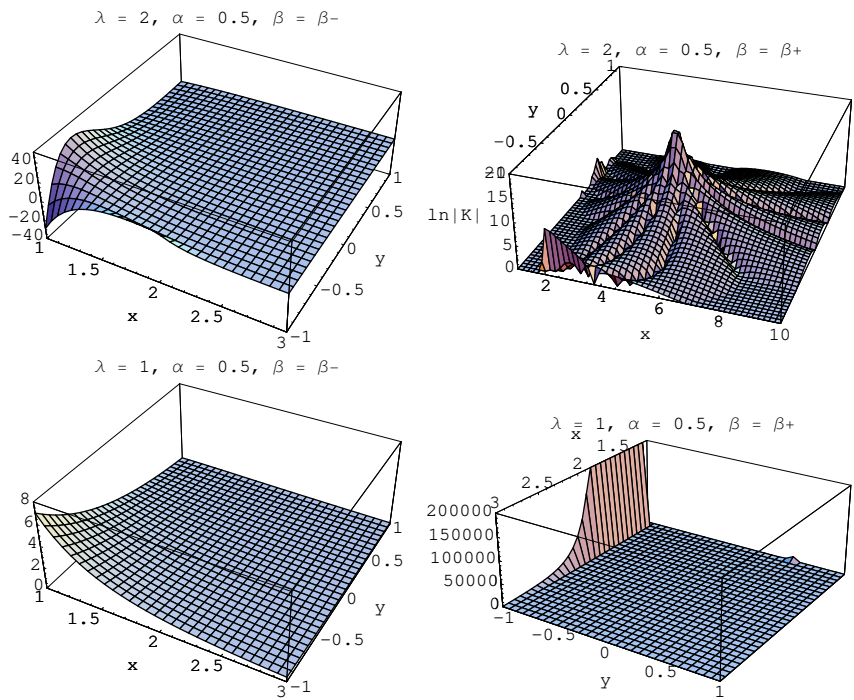

  \includegraphics[scale=0.7]{KI_l2_r.epsi}
  \includegraphics[scale=0.7]{KI_l2_s.epsi}\\
  \includegraphics[scale=0.7]{KI_l1_r.epsi}
  \includegraphics[scale=0.7]{KI_l1_s.epsi}
  \caption{ Plots of Kretchman Invariants for the cases of 
  $(\lambda,\alpha,\beta,)=(2,0.5,\beta_{-})$,  
  $(2,0.5,\beta{+})$,  
  $(1,0.5, \beta{+})$, and  
  $(1,0.5,\beta{-})$.
   In the right-upper panel we plot the value of $\ln |K|$.
   We can see that $K$ grows disastrously around the point where $B=0$.
   }
 \label{fig:KI}
 \end{figure*}

The rod structure analysis shows that the solution
satisfies the necessary condition for the absence of the
curvature singularity on the $z$-axis.
It was shown in the above, however, that the function $B$ 
becomes zero for some cases.
The curvature singularity appears at the point where $B=0$.
We plot $K$ for the four representative cases 
in FIG. \ref{fig:KI}.
The left-upper panel corresponds to the case of $\beta=\beta_{-}$ and $\lambda>1$.
There cannot be seen a curvature singularity in this plot.
The right-upper panel corresponds to the case of $\beta=\beta_{+}$ and $\lambda>1$.
We can see that the value of curvature grows disastrously around
the point where $B=0$ in this plot.

When $\lambda=1$ there appears a directional curvature singularity.
For the singular case $\beta \ne \beta_{-} =0$,
the Kretchman invariant diverges at $x=1$ as
\begin{equation}
 K \propto \frac{72 \beta^4}{(1+\alpha\beta)^4 \sigma^2 (x-1)^4}
\end{equation}
along the plane $y=-1$. 
While this value is finite when we approach $y=-1$ on the event
horizon $x=1$,
\begin{equation}
 K  =  \frac{9(1+\alpha\beta)^4}{2\beta^4\sigma^2} .
\end{equation}
The right-lower panel of FIG. \ref{fig:KI}
shows the behavior $K$ of this case.
For the regular case $\beta = \beta_{-} =0$,
there does not appear curvature singularity at $x=1$ and $y=-1$
and the value of $K$ is obtained as
\begin{equation}
 K = \frac{9(1+\alpha^2)^2}{2\sigma^2}.
\end{equation}
The left-lower panel of FIG. \ref{fig:KI}
shows the behavior $K$ of single-rotational black hole.


\subsection{C-metric expression}
\label{sec:C-metric}
The metric  (\ref{full_metric}) is rewritten by the C-metric coordinates
\cite{Figueras:2005zp} as
\begin{eqnarray}
 ds^2 &=& -\frac{H(\lambda_c,v,u)}{H(\lambda_c,u,v)} 
 \left[dx^0 - \frac{\lambda_c a_c v (1-u^2)}{H(\lambda_c,v,u)}
  d\bar{\phi}\right]^2 \nonumber \\
 && + \frac{R_c^2}{(u - v)^2} H(\lambda_c,u,v)
 \left[ -\frac{dv^2}{(1-v^2)F(\lambda_c,v)} 
 \right. \nonumber \\ && \left.
 - \frac{(1-v^2)F(\lambda_c,u)}{H(\lambda_c,u,v)} d\bar{\psi}^2 
  + \frac{du^2}{(1-u^2)F(\lambda_c,u)} 
 \right. \nonumber \\ && \left.
  + \frac{(1-u^2)F(\lambda_c,v)}{H(\lambda_c,v,u)} d\bar{\phi}^2\right],
  \label{C-metric}
\end{eqnarray}
where
\begin{eqnarray}
&& F(\lambda_c,\xi) = 1 + \lambda_c \xi + \left(\frac{a_c \xi}{R_c}\right)^2, \nonumber \\ 
&& H(\lambda_c,\xi_1,\xi_2) = 1 + \lambda_c \xi_1 
                    + \left(\frac{a_c \xi_1 \xi_2}{R_c }\right)^2,
\end{eqnarray}
and $-1<u<1$ and $-\infty < v <-1$.
Here we give the relation between the prolate-spheroidal coordinates $(x,y)$
and the C-metric coordinates $(u,v)$. 
After a rather lengthy calculation, the metric (\ref{full_metric})
written by the prolate-spheroidal coordinates  
can be transformed into the expression (\ref{C-metric})
by using the following coordinate transformations,
\begin{eqnarray}
 x &=& \frac{2(u-v_i)(v-v_h)}{(u-v)(v_h-v_i)} +1\\
 y &=& \frac{uv - 1}{u - v}, \\
 \phi &=& \frac{a_c\sqrt{(1+v_h)(1+v_i)}}{R_c}  \bar{\phi} , \\
 \psi &=& \frac{a_c\sqrt{(1+v_h)(1+v_i)}}{R_c}  \bar{\psi},
\end{eqnarray}
where
\begin{eqnarray}
v_h &=& \frac{R_c^2}{2 a_c^2}\left[-\lambda_c + \sqrt{\lambda_c^2 - \frac{4a_c^2}{R_c^2} }\right] \\
v_i &=& \frac{R_c^2}{2 a_c^2}\left[-\lambda_c - \sqrt{\lambda_c^2 - \frac{4a_c^2}{R_c^2} }\right].
\end{eqnarray}
In addition the relation between the parameters $(\sigma,\lambda,\alpha,\beta)$
and $(R_c,\lambda_c,a_c)$,
\begin{eqnarray}
\sigma &=& \frac{R_c^2 (v_h - v_i)}{2(1+v_h)(1+v_i)} \\
\lambda &=& \frac{v_h v_i -1}{v_h -v_i}\\
\alpha &=& -\frac{\sqrt{2(v_h -v_i)}}{v_i + 1}\\
\beta &=& \frac{v_h + 1}{\sqrt{2(v_h -v_i)}},
\end{eqnarray}
should be used there.  We can easily confirm that the no CTC 
condition $\beta=\beta_-$ is satisfied.
Also, when the event horizon degenerates, $v_h=v_i$,
the parameters show the same behavior 
as the case of maximum rotation Eq. (\ref{eq:para_extreme}).
Inversely the parameters $\lambda_c$, $R_c^2$ and $a_c^2$ can be written as
\begin{eqnarray}
\lambda_c &=& \frac{\lambda - 1 -(\lambda + 1) \alpha^2 \beta^2}
       {(\lambda(1+\alpha\beta) - 1)(\lambda(1+\alpha\beta) + \alpha\beta)}
        \nonumber\\
       &=& \frac{2\sqrt{\lambda^2-1}(1-h^2)}
      {(\sqrt{\lambda-1}+h\sqrt{\lambda+1})(\sqrt{\lambda+1}+h\sqrt{\lambda-1})} \\
    R_c^2 &=& -\frac{4\sigma \beta}{\alpha} \nonumber\\
          &=& \frac{m^2 h^2}{1-h^2} \label{eq:Rc} \\
    a_c^2 &=& \frac{4\sigma \beta^2}
{(\lambda(1+\alpha\beta) - 1)(\lambda(1+\alpha\beta) + \alpha\beta)} \nonumber \\
          &=& \frac{m^2 h^2}{1-h^2} 
  \frac{(\sqrt{\lambda-1}-h\sqrt{\lambda+1})(\sqrt{\lambda+1}-h\sqrt{\lambda-1})}
       {(\sqrt{\lambda-1}+h\sqrt{\lambda+1})(\sqrt{\lambda+1}+h\sqrt{\lambda-1})}. \nonumber \\
\end{eqnarray}
By using Eqs. (\ref{def:h}) and (\ref{eq:Rc}) we can rewrite Eq. (\ref{in_phi}) as
\begin{equation}
 \Delta \phi = 2\pi h = \frac{2\pi}{\sqrt{1+\frac{m^2}{R_c^2}}}.
\end{equation}
Therefore the period of $\phi$ is determined only by the ratio of
 the mass parameter $m$ to the radius parameter $R_c$.

\begin{figure*}
  \includegraphics[scale=0.4,angle=270]{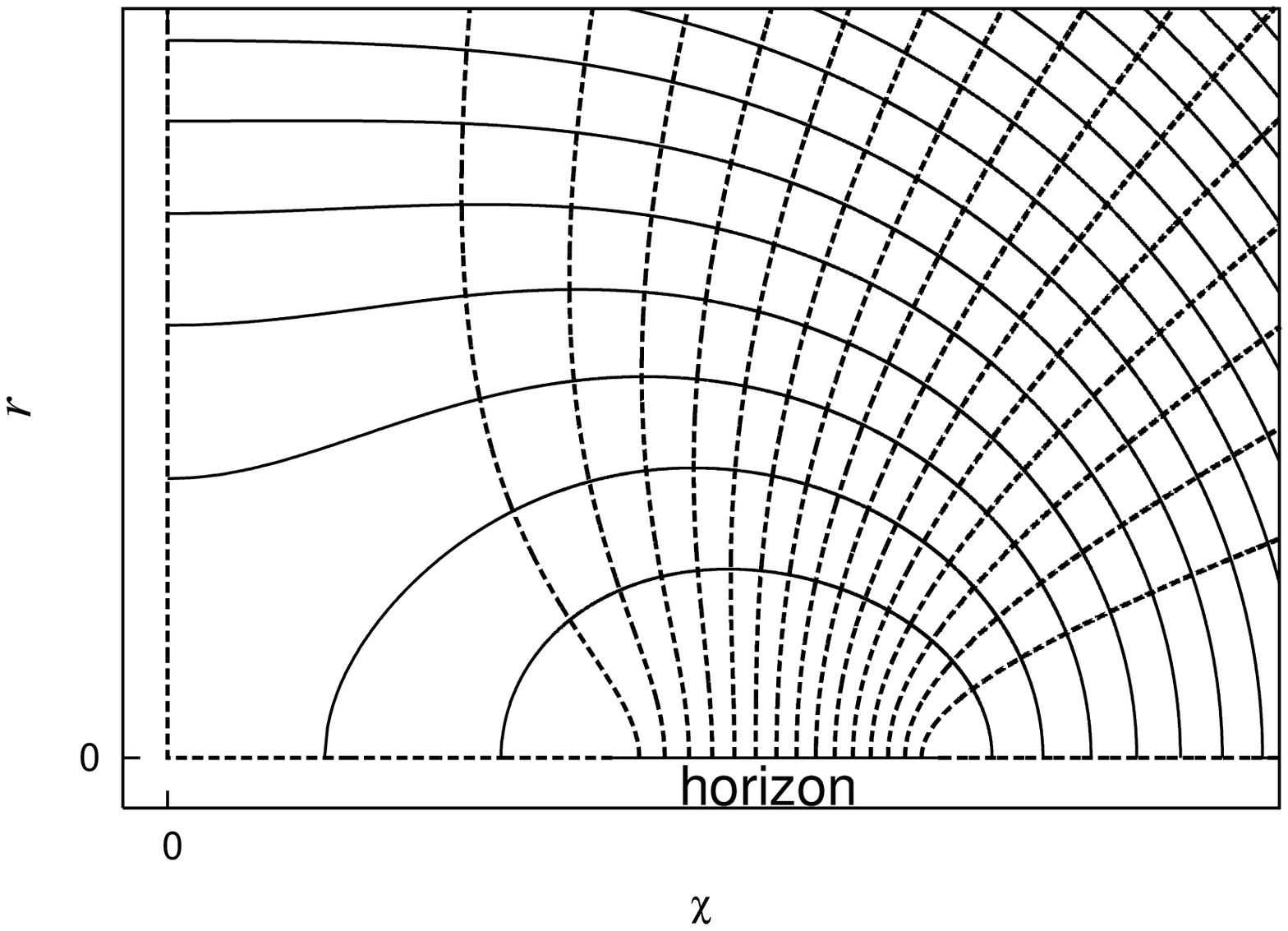}
  \includegraphics[scale=0.4,angle=270]{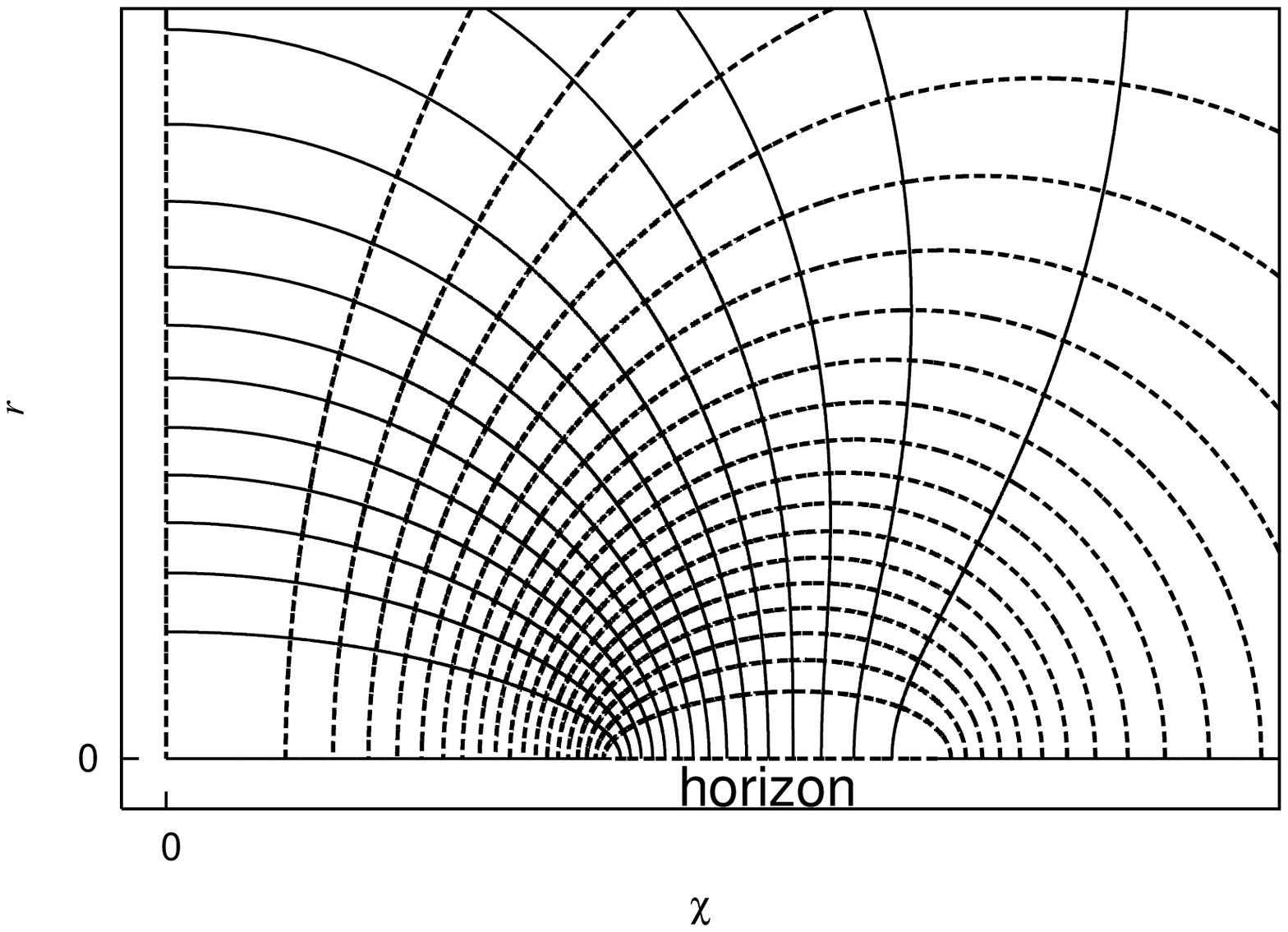}
  \caption{ 
  The contour lines of $x$, $y$, $u$ and $v$ in $\chi-r$ plane.
  In the left panel, the solid lines and the dotted lines are $x=$ const. 
  and $y =$ const. lines respectively. 
  In the right panel, the solid lines and the dotted lines are $u=$ const. 
  and $v =$ const. lines respectively. }
 \label{fig:coord}
 \end{figure*}

In FIG. \ref{fig:coord} we present schematic pictures for 
the relations between $(x,y)$ and $(r,\chi)$ coordinates and 
between  $(u,v)$ and $(r,\chi)$ coordinates.
Here we use the relations 
\begin{eqnarray}
 && \hspace{-0.5cm}\chi^2 = \sigma (\sqrt{(x^2-1)(1-y^2)+(xy+\lambda)^2} + xy+\lambda), ~~~\\
 && \hspace{-0.5cm}r^2 = \sigma (\sqrt{(x^2-1)(1-y^2)+(xy+\lambda)^2} - xy-\lambda).
\end{eqnarray}
The lines of $u=$const. and $v=$const. are denser than
$x$ and $y$ near the inner edge of the  horizon.


\section{summary}
\label{sec:summary}
In this paper we have described the solution-generating technique 
for the Einstein equation of five-dimensional General Relativity. 
Using this method we can systematically construct axisymmetric stationary
solutions with asymptotic flatness. 
If we prepare various seed solutions, we can obtain different
kinds of new solutions with the single-rotation.
In our analysis we adopted the procedure given by Castejon-Amenedo and Manko
to derive the exact form of the metric functions. 

For the application of the method, we adopted the
Minkowski spacetime as the simplest seed solution. 
Although the seed solution we adopted is so simple, 
the obtained series of the solutions has scientific importance. 
It includes two important limits, the Myers and Perry 
single-rotational black hole
and the rotational black string. 
More significantly, the part of the series 
should be one single-rotational limit
of an undiscovered double-rotational black ring 
which has the Emparan-Reall's black ring as another limit.

We have examined the qualitative features of the solutions in detail.
Generally, there are ergo-regions 
around the local black objects because of the rotation.
In addition, there exist regions where
the metric function $g_{\phi\phi}$ becomes negative.
In these regions closed timelike curves can be exist.  
However, we have confirmed
that there is no CTC region when the condition $\beta=\beta_-$ holds.
Even in this case, the conical singularity is inevitable inside or outside
the ring. Therefore we need the strut structure inside the ring 
because the periods of $\phi$ inside the ring is always smaller than
those of $\phi$ outside the ring and of $\psi$.
In addition we have shown that there is an upper limit of the rotation
parameter $a_0$ when we fix the mass parameter $m$ and the radius parameter 
$R_c$.
When the condition $\beta=\beta_+$ holds, we can also define the period
of $\phi$ inside the ring. In this case there is positive
deficit angles inside the ring in contrast to the case of $\beta=\beta_-$.
We have investigated the behavior of the curvature invariant.
This variable is finite when the condition $\beta=\beta_-$ holds and
can diverge where $B=0$ for another cases. 
We have also shown that there is a directional 
curvature singularity  for the case of $\lambda=1$.

By the rod structure analysis we have understood the relation between
the seed and the obtained solutions. By analogy of this relation
we can obtain a seed of the $S^1$-rotating black ring solution 
\cite{Iguchi:2006rd}.

Finally, we derived the relations between the prolate-spheroidal coordinates
and the C-metric coordinates which was derived by Figueras
\cite{Figueras:2005zp}.
We confirmed the equivalence between the metric (\ref{full_metric}) 
with the condition $\beta=\beta_-$ 
and the expression given by Figueras.

In the method presented here we can also adopt other seed 
spacetimes, so that 
we can generate some new solutions.
Although the solution obtained here has some pathologies including 
inevitable one, i.e., conical singularities, we can expect to obtain
the new solution without these pathologies by an adequate seed metric
as in the case of the $S^1$-rotating black ring which is reduced from
the seed of Wick rotated C-metric solution \cite{Iguchi:2006rd}.
However it should be noticed that the method introduced here cannot be 
used for the solution generation of double-rotational black rings
because of the metric form (\ref{WPmetric}). 
For this purpose other methods may be used.
One of the powerful tools would be the inverse scattering 
method \cite{refBZ}. 

\acknowledgments
This work is partially supported by Grant-in-Aid for Young Scientists (B)
(No. 17740152) from Japanese Ministry of Education, Science,
Sports, and Culture
and by Nihon University Individual Research Grant for
2005.

\appendix
\section{Neugebauer and Kramer representation}
\label{app:NK}
To derive the potential functions $a$ and $b$, it is convenient to use 
the Neugebauer and Kramer's powerfull representation for Ernst potential \cite{Stephani:2003tm}.
Adding the $2N$ solitons to a static seed potential ${\cal E}^{(0)}$, 
the new Ernst potential is given by
\begin{equation}
{\cal E}={{\cal E}^{(0)}}
         \frac{\det\left(\frac{\al_p R_p-\al_q R_q}{K_p-K_q}-
                      1\right)}
              {\det\left(\frac{\al_p R_p-\al_q R_q}{K_p-K_q}+1\right)}
 \label{Ernst_NK1}
\end{equation}
where $p=1,3,\dots,2N-1$, $q=2,4,\dots,2N$ and $R_{j}=\sqrt{\rho^2+(z-K_j)^2}$. 
The function $\al_j$ is given by 
\begin{equation}
\al_j=\frac{l_j+ie^{2\phi_j}}{l_j-ie^{2\phi_j}}\,.
\end{equation}
Here the quantity $l_j$ is an integral constant, and 
the function $\phi_{j}$ obeys the following Riccati equations,
\begin{equation}
d\phi_j=\frac{1}{2}\left[\,\,
    \left(\gamma_j\right)^{1/2}\del{\zeta}\ln {\cal E}^{(0)} d\zeta
                  +\left(\gamma_j\right)^{-1/2}\del{\bar{\zeta}}
                                         \ln {\cal E}^{(0)} d\bar{\zeta}
                   \,\,\right]\,, \label{diff_phi}
\end{equation}
where $\zeta=\rho+iz$, and $\gamma_j=(K_j-i\bar{\zeta})/(K_j+i\zeta)$.
For the case of 
$\ln {\cal E}^{(0)}=\frac{1}{2}\ln\left[\,R_{d}+(z-d)\,\right]$, 
the function $\phi_j$ is given by 
\begin{equation}
\phi_j=\phi_{d,j}=\frac{1}{2}
 \ln\left[\,e^{-\tilU_{d}}\left(e^{2U_j}+e^{2\tilU_{d}}\right)\,\right],
\end{equation}
where $\tilU_{d}:=\frac{1}{2}\ln\left[\,R_{d}+(z-d)\,\right]$ and 
$U_{j}:=\frac{1}{2}\ln\left[\,R_{j}-(z-K_j)\,\right]$. 

Next we consider the relation between the Ernst potentials (\ref{Ernst_GM}) and (\ref{Ernst_NK1}).
The parameter $K_1$ and $K_2$ are set to $\sigma$ and $-\sigma$, respectively 
so that 
the useful expression,
\begin{equation}
R_{\sigma}=\sigma(x-y)\, \ \ \ R_{-\sigma}=\sigma(x+y)
\end{equation}
are derived.
Then 
the equation (\ref{Ernst_NK1}) becomes
\begin{eqnarray}
{\cal E}_{S}
&=&e^{S^{(0)}}
 \frac{(x-y)\al_{\sigma}-(x+y)\al_{-\sigma}-2}
      {(x-y)\al_{\sigma}-(x+y)\al_{-\sigma}+2}\ , \label{Ernst_NK2}
\end{eqnarray}
where $\al_{\sigma}=\al_{1},\ \al_{-\sigma}=\al_{2}$.
When we rewrite Eq. (\ref{Ernst_GM}), we obtain the following expression for the Ernst potential,
\begin{eqnarray}
{\cal E}_{S}&=&
 e^{S^{(0)}}
 \frac{(x-y){\displaystyle \frac{1+ia}{1-ia}}
             +(x+y){\displaystyle \frac{1+ib}{1-ib}-2}}
     {(x-y){\displaystyle \frac{1+ia}{1-ia}}
             +(x+y){\displaystyle \frac{1+ib}{1-ib}+2}}\ . \label{Ernst_GM2}
\end{eqnarray}
Comparing Eqs. (\ref{Ernst_NK2}) and (\ref{Ernst_GM2}), we obtain the following relations,
\begin{equation}
a=i\frac{1-\al_{\sigma}}{1+\al_{\sigma}},
  \ \ \ b=i\frac{1+\al_{-\sigma}}{1-\al_{-\sigma}}\ .
\end{equation}
Here the functions $\al_{\sigma}$ and $\al_{-\sigma}$ are related to 
 $\phi_{\sigma}$ and $\phi_{-\sigma}$, respectively, 
through the equations 
 \begin{equation}
 \al_{\sigma}
  =\frac{l_{\sigma}+ie^{2\phi_{\sigma}}}{l_{\sigma}-ie^{2\phi_{\sigma}}}\ ,\ \ 
\al_{-\sigma}
  =\frac{l_{-\sigma}+ie^{2\phi_{-\sigma}}}{l_{-\sigma}-ie^{2\phi_{-\sigma}}}\ ,
\end{equation}
so that $a$ and $b$ are expressed with $\phi_{\sigma}$ and $\phi_{-\sigma}$, 
\begin{equation}
a=l_{\sigma}^{-1}e^{2\phi_{\sigma}}\ ,\ \ \ 
b=-l_{-\sigma}e^{-2\phi_{-\sigma}}\ . 
\end{equation}
Hence the functions $a$ and $b$, and the corresponding full metric 
can be determined, once the functions $\phi_{\sigma}$ and $\phi_{-\sigma}$ 
are derived.
\section{metric functions in canonical coordinates}
\label{app:canonical}
Here we rewrite the metric (\ref{full_metric}) in the canonical 
coordinates $\rho$ and $z$.
At first, the last term of Eq. (\ref{full_metric}) can be written 
into the following form,
\begin{eqnarray} 
ds^2 &=&-\frac{A}{B}\left[dx^0-\left(2\sigma e^{-S^{(0)}}
         \frac{C}{A}+C_1\right) d\phi\right]^2    \nonumber  \\ &&
+\frac{B}{A}e^{-2S^{(0)}}\rho^2(d\phi)^2+e^{2S^{(0)}}(d\psi)^2
                  \nonumber \\ && 
+\frac{C_2 \sigma^2}
{\sqrt{2}(1+\lambda)}\frac{B}
{R_{\sigma}R_{-\sigma}R_{-\lambda\sigma}}
\nonumber  \\ && \times
 \frac{(\lambda-1)R_{\sigma}+(\lambda+1)R_{-\sigma}+2R_{-\lambda\sigma}}
      {R_{\sigma}+R_{-\sigma}+2\sigma}\left(d\rho^2+dz^2\right), \nonumber \\
\label{canonical_metric}
\end{eqnarray}
by using the definition of $R_d$.
To obtain the expressions of $A$, $B$ and $C$, 
we use the following relations
\begin{eqnarray}
 x+1 &=& \frac{e^{2\tU_{-\sigma}}+e^{2U_{\sigma}}}{2\sigma}
    =\frac{R_\sigma+R_{-\sigma}+2\sigma}{2\sigma},  \\
 x-1 &=& \frac{e^{2\tU_{\sigma}}+e^{2U_{-\sigma}}}{2\sigma}
    =\frac{R_\sigma+R_{-\sigma}-2\sigma}{2\sigma}, \\
 1+y &=& \frac{e^{2\tU_{-\sigma}}-e^{2\tU_{\sigma}}}{2\sigma}
    =\frac{R_{-\sigma}-R_{\sigma}+2\sigma}{2\sigma}, \\
 1-y &=& \frac{e^{2U_{\sigma}}-e^{2U_{-\sigma}}}{2\sigma}
    =\frac{R_\sigma-R_{-\sigma}+2\sigma}{2\sigma}.
\end{eqnarray}
In addition the forms of the potential functions (\ref{eq:a_pot}) and
(\ref{eq:b_pot}) are written as
\begin{eqnarray}
 a &=& \frac{\alpha}{2\sigma^{1/2}}
 \frac{e^{2U_{\sigma}}+e^{2\tU_{-\lam\sigma}}}{e^{\tU_{-\lam\sigma}}} \nonumber \\
 &=&  \frac{\alpha}{2\sigma^{1/2}}
    \frac{R_\sigma+R_{-\lam\sigma}+(\lam+1)\sigma}
         {\sqrt{R_{-\lam\sigma}+(z+\lam\sigma)}}, \\
 b &=& 2\sigma^{1/2}\beta
 \frac{e^{\tU_{-\lam\sigma}}}{e^{2U_{-\sigma}}+e^{2\tU_{-\lam\sigma}}}  \nonumber \\
   &=&  2\sigma^{1/2}\beta
    \frac{\sqrt{R_{-\lam\sigma}+(z+\lam\sigma)}}
         {R_{-\sigma}+R_{-\lam\sigma}+(\lam-1)\sigma}.
\end{eqnarray}
Using these functions, we obtained the functions
$A$, $B$ and $C$ in the canonical coordinates as
\begin{widetext}
\begin{eqnarray*}
 A &=& \frac{1}{(2\sigma)^2}
  \left\{\left(e^{2\tU_{-\sigma}}+e^{2U_{\sigma}}\right)
    \left(e^{2\tU_{\sigma}}+e^{2U_{-\sigma}}\right) (1+ab)^2 
        -\left(e^{2\tU_{-\sigma}}-e^{2\tU_{\sigma}}\right)
    \left(e^{2U_{\sigma}}-e^{2U_{-\sigma}}\right) (b-a)^2\right\} \\
   &=& \frac{1}{(2\sigma)^2}  
       \left\{
        \left(\left(R_{\sigma}+R_{-\sigma}\right)^2-(2\sigma)^2\right)
              (1+ab)^2
       +\left(\left(R_{\sigma}-R_{-\sigma}\right)^2-(2\sigma)^2\right)
              (b-a)^2
       \right\}, \\
 B &=& \frac{1}{(2\sigma)^2} 
  \left\{\left[\left(e^{2\tU_{-\sigma}}+e^{2U_{\sigma}}\right)
    +\left(e^{2\tU_{\sigma}}+e^{2U_{-\sigma}}\right)ab \right]^2
    +\left[\left(e^{2\tU_{-\sigma}}-e^{2\tU_{\sigma}}\right)a
    -\left(e^{2U_{\sigma}}-e^{2U_{-\sigma}}\right)b\right]^2\right\}
 \\
   &=& \frac{1}{(2\sigma)^2} 
     \left\{
       \left[
         \left(R_{\sigma}+R_{-\sigma}+2\sigma\right)
       + \left(R_{\sigma}+R_{-\sigma}-2\sigma\right) a b
       \right]^2
      +\left[
         \left(R_{\sigma}-R_{-\sigma}+2\sigma\right)b
       - \left(R_{\sigma}-R_{-\sigma}-2\sigma\right)a
       \right]^2
     \right\}, \\
 C &=& \frac{1}{(2\sigma)^3} 
     \left\{
         \left(e^{2\tU_{-\sigma}}+e^{2U_{\sigma}}\right)
         \left(e^{2\tU_{\sigma}}+e^{2U_{-\sigma}}\right) (1+ab)
       \left(  \left(e^{2U_{\sigma}}-e^{2U_{-\sigma}}\right) b
      -\left(e^{2\tU_{-\sigma}}-e^{2\tU_{\sigma}}\right) a \right)
     \right. \nonumber \\  && \left. \hspace{0.cm}
     +\left(e^{2\tU_{-\sigma}}-e^{2\tU_{\sigma}}\right)
       \left(e^{2U_{\sigma}}-e^{2U_{-\sigma}}\right) (b-a)
      \left(\left(e^{2\tU_{-\sigma}}+e^{2U_{\sigma}}\right)
        - \left(e^{2\tU_{\sigma}}+e^{2U_{-\sigma}}\right) ab\right)
     \right\} \\
   &=& \frac{1}{(2\sigma)^3}
      \left\{
       \left(\left(R_{\sigma}+R_{-\sigma}\right)^2-(2\sigma)^2\right)
              (1+ab)
      \left(\left(R_{\sigma}-R_{-\sigma}+2\sigma\right) b
           +\left(R_{\sigma}-R_{-\sigma}-2\sigma\right) a \right)
      \right. \nonumber \\  && \left. \hspace{0cm}
    - \left(\left(R_{\sigma}-R_{-\sigma}\right)^2-(2\sigma)^2\right)
            (b-a)
      \left(\left(R_{\sigma}+R_{-\sigma}+2\sigma\right) 
           -\left(R_{\sigma}+R_{-\sigma}-2\sigma\right) a b\right)
      \right\} .
\end{eqnarray*}

From these results, the metric which corresponds to the static case $a=b=0$ 
is reduced to 
\begin{eqnarray}
ds^2
&=& 
-\frac{R_{\sigma}+z-\sigma}{R_{-\sigma}+z+\sigma}(dx^0)^2
+\frac{(R_{-\sigma}+z+\sigma)(R_{-\lambda\sigma}-z-\lambda\sigma)}
 {R_{\sigma}+z-\sigma}(d\phi)^2
+\left(R_{-\lambda\sigma}+z+\lambda\sigma\right)(d\psi)^2   \nonumber\\
&& +\frac{C_2}{\sqrt{2}(\lambda+1)}
 \frac{(R_{\sigma}+R_{-\sigma}+2\sigma)
       [(\lambda-1)R_{\sigma}+(\lambda+1)R_{-\sigma}+2R_{-\lambda\sigma}]}
 {R_{\sigma}R_{-\sigma}R_{-\lambda\sigma}}
      \left(d\rho^2+dz^2\right)\,.
\label{eq:static_cano}
\end{eqnarray}
\end{widetext}
Here we used the following relations,
\begin{eqnarray}
&& \frac{R_{\sigma}+R_{-\sigma}-2\sigma}{R_{\sigma}+R_{-\sigma}+2\sigma}
=\frac{R_{\sigma}+z-\sigma}{R_{-\sigma}+z+\sigma}, \\
&& \rho^2=( R_{-\lambda\sigma}+z+\lambda\sigma )
       ( R_{-\lambda\sigma}-z-\lambda\sigma )\,.
\end{eqnarray}
After a trivial coordinate transformation and change of parameters, 
we see that the metric form is equivalent to the form which was 
described in \cite{refHAR}.

\section{Direct generation of limit solutions}
\label{app:direct}
The limit solutions (\ref{eq:MPmetric_1}) and the rotational black string
are obtained directly from the corresponding seed solutions.
\subsection{single-rotational Myers and Perry black hole}
The seed metric of the Myers and Perry black hole with a single-rotation
can be derived from Eq. (\ref{eq:metirc_M}) with $\lambda=1$,
\begin{eqnarray}
ds^2 &=&
\,-(dx^0)^2+\left(\sqrt{\rho^2+(z+\sigma)^2}-(z+\sigma) \right)d\phi^2
     \nonumber \\ &&
  +\left(\sqrt{\rho^2+(z+\sigma)^2}+(z+\sigma) \right)d\psi^2
   \nonumber \\
 && +\frac{1}{2\,\sqrt{\rho^2+(z+\sigma)^2}}(d\rho^2+dz^2). \label{seed_MP}
\end{eqnarray}
We can read out the seed functions from the above metric as
\begin{equation}
S^{(0)}=T^{(0)}=
\frac{1}{2}\ln
      \left[\sqrt{\rho^2+(z+\sigma)^2}+(z+\sigma)\right] 
      =\tilU_{- \sigma}.
\end{equation}
Using Eqs. (\ref{ab_phi}) and (\ref{phi_i}) 
we obtain the functions $a$ and $b$,  
\begin{equation}
a = \al\,\, \sqrt{\frac{x+1}{1+y}}, ~~~
b = \beta\,\, \frac{\sqrt{(x+1)(1+y)}}{x+y}.
\end{equation}
The solution of the differential equations (\ref{eq:drho_gamma'}) 
and (\ref{eq:dz_gamma'}) can be obtained as
\begin{equation}
 e^{2\gamma'} = \frac{(x^2-1)(1+y)}{\sqrt{2}(x^2-y^2)}.
\end{equation}
Using these results and the no CTC condition $\beta=0$ we can rederive
the metric (\ref{eq:MPmetric_1}).

\subsection{rotational black string}
The seed metric of the rotational black string solution is
\begin{equation}
ds^2 = -(dx^0)^2+ \frac{\rho^2}{\sigma} d\phi^2 + d\rho^2 + dz^2 
+ \sigma d\psi^2. \label{seed_Kerr}
\end{equation}
The corresponding seed functions become
\begin{equation}
 S^{(0)} = T^{(0)} = \frac{1}{2} \ln \sigma.
\end{equation}
The functions $a$ and $b$ can be obtained as 
\begin{equation}
 a=\alpha, ~~~ b=\beta,
 \label{eq:ab_BS}
\end{equation}
trivially. The corresponding $\gamma'$ becomes
\begin{equation}
 e^{2\gamma'} = \frac{x^2-1}{x^2-y^2}.
 \label{eq:gamd_BS}
\end{equation}
As a result, we can derive the corresponding metric form
\begin{eqnarray}
 ds^2 &=& - \frac{A}{B}\left(dx^0 
 -\left(2\sigma^{\frac{1}{2}}\frac{C}{A}+C_1 \right) d\phi \right)^2 
 + \frac{B}{A} \frac{\rho^2}{\sigma}d\phi^2  \nonumber \\
&& +C_2 B\left(\frac{dx^2}{x^2 -1}+\frac{dy^2}{1-y^2}\right)
 +\sigma d\psi^2 ,
\end{eqnarray}
where $A$, $B$ and $C$ are obtained by replacing $a$ and $b$ to $\alpha$ and $\beta$ in Eqs. (\ref{eq:A}),  (\ref{eq:B}) and (\ref{eq:C}), respectively. 
The four-dimensional part of this solution corresponds with the Kerr-NUT solution \cite{Demianski:1966}.
In fact the complex potential $\xi \equiv (1- {\cal E}_S)/(1+ {\cal E}_S)$ can be represented as
\begin{equation}
 \xi^{-1} 
          = \frac{1+\alpha\beta}{(1-i\alpha)(1-i\beta)}x
             -i\frac{\alpha-\beta}{(1-i\alpha)(1-i\beta)}y
\end{equation}
and then we can confirm that the Kerr and NUT solutions correspond
with the cases $\alpha=-\beta$ and
$\alpha=\beta$, respectively. 
Comparing this and the expression written by the Kerr and NUT parameters
$\theta$ and $\varphi$,
\begin{equation}
 \xi^{-1}  = e^{i\varphi}(\cos \theta x - i \sin \theta y),
\end{equation}
we can obtain the following relations between the parameters
\begin{equation}
 \tan \theta = \frac{\alpha-\beta}{1+\alpha\beta},~~~ \mbox{and}~~~
 \tan(2\varphi) = \frac{2(\alpha+\beta)(\alpha\beta-1)}
                {(\alpha+\beta)^2-(\alpha\beta-1)^2}.
\end{equation}
Therefore we can rederive the metric (\ref{eq:metric_BS}) by using 
Eqs. (\ref{eq:ab_BS}) and (\ref{eq:gamd_BS}) 
with the condition $\alpha=-\beta$.

\section{rod structure analysis}
\label{app:rod}
In this appendix we give a brief explanation of the
rod structure analysis erabolated by Harmark \cite{refHAR}.
See \cite{refHAR} for complete explanations.

Here we denote the D-dimensional axially symmetric 
stationary metric as
\begin{equation}
 ds^2 = G_{ij}dx^i dy^j + e^{\nu}(d\rho^2 + dz^2)
\end{equation}
where $G_{ij}$ and $\nu$ are functions only of $\rho$ and $z$
and $i,j=0,1,\dots,D-3$. 
The $D-2$ by $D-2$ matrix field $G$ satisfies the following
constraint
\begin{equation}
 \rho = \sqrt{|\det G|}.
\end{equation}
The equations for the matrix field $G$ 
can be derived from the Einstein equation $R_{ij}=0$ as
\begin{equation}
 G^{-1}\nabla G = (G^{-1} \nabla G)^2,
 \label{eq:har}
\end{equation}
where the differential operator $\nabla$ is 
the gradient in three-dimensional unphysical flat space with metric
\begin{equation}
 d\rho^2 + \rho^2 d\omega^2 + dz^2.
\end{equation}

Because of the constraint $\rho = \sqrt{|\det G|}$,
at least one eigenvalue of $G(\rho,z)$ goes to zero for 
$\rho \rightarrow 0$. However it was shown that
if more than one eigenvalue goes to zero as $\rho \rightarrow 0$,
we have a curvature singularity there. Therefore  we consider
solutions which have only one eigenvalue goes to zero for 
$\rho \rightarrow 0$, except at isolated values of $z$.
Denoting these isolated values of $z$ as $a_1,a_2,\dots,a_N$,
we can divide the $z$-axis into the $N+1$ intervals
$[-\infty,a_1]$,$[a_1,a_2]$,$\dots$,$[a_N,\infty]$,
which is called as rods. These rods correspond to the source
added to the equation (\ref{eq:har}) at $\rho=0$ to
prevent the break down of the equation there.

The eigenvector for the zero eigenvalue of $G(0,z)$
\begin{equation}
 {\bf v}=v^i \frac{\partial}{\partial x^i}
\end{equation}
which satisfies
\begin{equation}
 G_{ij}(0,z) v^i = 0,
\end{equation}
determines the direction of the rod.
If the value of $\frac{G_{ij}v^i v^j}{\rho^2}$
is negative (positive) for $\rho \rightarrow 0$ the rod is called
timelike (spacelike).
Each rod corresponds to the region of the translational or
rotational invariance
of its direction.
The timelike rod corresponds to a horizon.
The spacelike rod corresponds to a compact direction.
\end{document}